\documentclass[%
 reprint,
 superscriptaddress,
 amsmath,amssymb,
 aps,
prb,
]{revtex4-2}

\usepackage{graphicx}
\usepackage{caption}
\usepackage{float}
\usepackage{dcolumn}
\usepackage{bm}
\usepackage{bbold}
\usepackage{mathtools}
\usepackage{hyperref}  
\hypersetup{
    colorlinks=true,
    linkcolor=blue,
    filecolor=blue,      
    urlcolor=blue,
    citecolor=blue,
    }
\usepackage{xcolor}
\usepackage{caption}
\usepackage{multirow}

\newcommand{\kk}{\boldsymbol{k}}
\newcolumntype{C}{>{$\displaystyle}c<{$}}
\renewcommand{\a}{{\boldsymbol{a}}}
\renewcommand{\b}{{\boldsymbol{b}}}
\newcommand{\K}{{\boldsymbol{K}}}
\newcommand{\rr}{{\boldsymbol{r}}}

\makeatletter
\def\maketitle{
\@author@finish
\title@column\titleblock@produce
\suppressfloats[t]}
\makeatother

\begin{document}

\preprint{APS/123-QED}

\title{Flat-band projected {\it versus} fully atomistic twisted bilayer graphene}

\author{Miguel S\'anchez S\'anchez}
\email{miguel.sanchez@csic.es}
\affiliation{Insituto de Ciencia de Materiales de Madrid ICMM-CSIC. Madrid (Spain)}
\author{Tobias Stauber}
\email{tobias.stauber@csic.es}
\affiliation{Insituto de Ciencia de Materiales de Madrid ICMM-CSIC. Madrid (Spain)}


\begin{abstract}

We benchmark the recently proposed projection method \cite{sanchez25} for magic-angle twisted bilayer graphene (MATBG) across various symmetry-breaking phases at charge neutrality. The flat-band projected solutions agree well with the full tight-binding, with band structures and total energies differing by only a few meV. The projection to the flat bands is justified, owing to the increased gap to the remote bands in the normal state. Moreover, we employ a novel set of order parameters that allow us to visualize the wave functions locally in real space and quantify the breaking of various symmetries in the correlated phases. These order parameters are suitable for characterizing MATBG and generic honeycomb systems.

\end{abstract}

\maketitle

\section{INTRODUCTION}

Magic-angle twisted bilayer graphene (MATBG) stands as a cornerstone in the field of moiré materials \cite{Nuckolls2024, Mak2022, Kennes2021}, offering an experimentally tunable platform for the study of topology and strong interactions. As such, it has attracted significant interest, and both theoretical and experimental efforts have uncovered a rich phase diagram, featuring a plethora of correlated phases \cite{bernevig21, lian21, Bultinck20, vafek20,kwan21, Lu2019, Cao2018_2, Sharpe2019, Serlin2020, Pierce2021, Wu2021, Wong2020, Xie2019, Jiang2019, Kerelsky2019, Nuckolls2023, Nuckolls2020, Kerelsky2019, Zondiner2020, tomarken19, Uri2020, rai24, Datta2023, zhou24, Tian2023, Saito2020, Stepanov2020, Yankowitz2019, DiBattista2022, Cao2018, Cao2021, Dutta2025, Arora2020, chichinadze20, Wu18, lian19, Wu19, Chou24, Roy19, Dong23, wang24_2, gonzalez19, Long2024, Christos2023, putzer25, Khalaf2021, ingham2023quadraticdiracfermionscompetition, Rozen2021, Saito2021, zhang2025heavyfermionsmassrenormalization, cao2020, Jaoui2022,Polshyn2019, Xie2021, Morissette2023}. 

Different analytical models tackle the apparent contradiction between the Mott-like phenomena---such as the quantum-dot-like density \cite{Wong2020,Xie2019,Jiang2019,Choi2019,Nuckolls2023} and the extensive entropy at intermediate temperatures \cite{Rozen2021,Saito2021,zhang2025heavyfermionsmassrenormalization}---and the nontrivial topology of MATBG \cite{song21,zang2022realspacerepresentationtopological}---exemplified by the emergence of anomalous Hall effect \cite{Sharpe2019,Serlin2020} and fractional Chern insulators \cite{Xie2021}. 
These models include the Topological Heavy Fermion model \cite{song22}, which introduces a set of topologically trivial local moments plus a set of dispersive electrons carrying the topology, and the Nonlocal Moments model \cite{ledwith25}, which considers the flat Chern bands exhibiting concentrated charge and Berry curvature.

In an earlier work \cite{sanchez25}, the Authors addressed a different tension, in that case related to the construction of the low-energy models. Starting from an interacting tight-binding model for the $\pi$ electrons of MATBG, we obtained a mean-field, symmetry-preserving solution that renormalizes both the dispersion and the wave functions of the non-interacting bands. An effective theory was then constructed via the many-body projection onto the desired low-energy degrees of freedom of this renormalized normal state---this procedure is analogous to the Wilsonian renormalization in which high-energy modes are successively integrated out \cite{vafek20,huang25}. The resulting theory exactly reproduces the symmetry-preserving state and is free from double-counting ambiguities \cite{potasz21,faulstich23,liu21,xie20,Bultinck20}.

In this work, we benchmark the accuracy of the flat-band projection. In Section \ref{sec2}, we compare the results for several symmetry-breaking phases at charge neutrality within the fully atomistic (full tight-binding) and flat-band projected models. We find good agreement between both setups, demonstrating the ability of the flat-band model to reproduce the results of the full theory. 

Moreover, we present a set of local order parameters---equivalent to those first introduced in Ref. \cite{sanchez24_2}---that generalize the microscopic valley operator \cite{lopez20,colomes18}, and use them to analyze the properties of the symmetry-breaking phases in Section \ref{sec3}. These novel order parameters allow us to visualize the wave functions locally in real space and complement the information of analogous quantities in momentum space \cite{Bultinck20,liu21,kwan21,kwan2023electronphononcouplingcompetingkekule,Hofmann22,soejima20,wagner22,wang23} or in the heavy fermion basis \cite{wang2024electronphononcouplingtopological,shi2024moireopticalphononsdancing}.

Finally, in Section \ref{sec4} we discuss the implications of our results, the choice of interaction parameters and possible extensions of our work. 

\begin{figure*}
    \includegraphics[width=.65\linewidth]{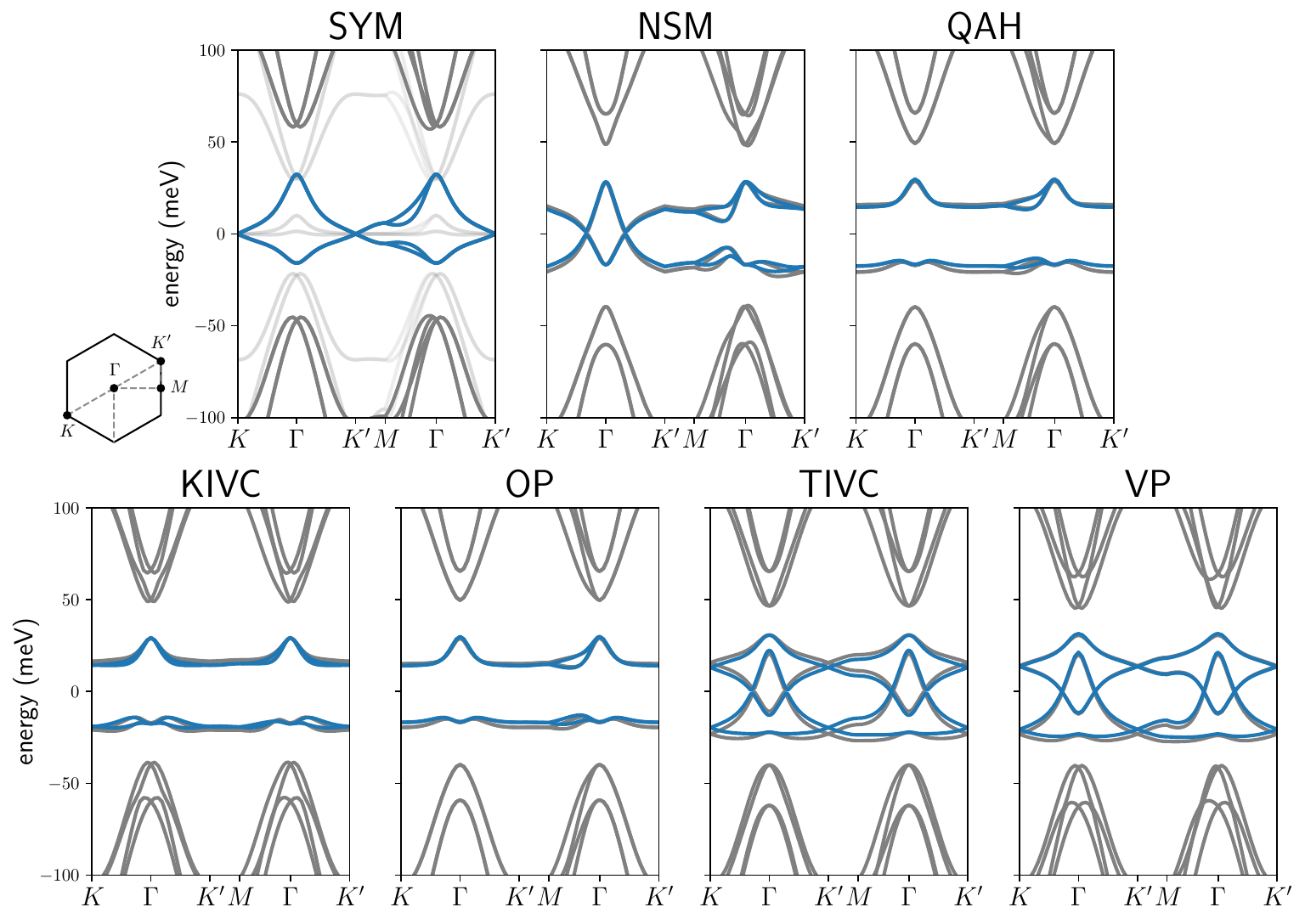}
    \caption{Band structures of the normal state (SYM), the nematic semimetal (NSM), quantum anomalous Hall (QAH), Kramers intervalley coherent (KIVC), orbital polarized (OP), time-reversal intervalley coherent (TIVC) and valley polarized (VP) states. Bands from the tight-binding and flat-band models are shown in gray and blue, respectively. We also plot the non-interacting bands in the SYM panel in light gray, for comparison. In the NSM, the protected Dirac nodes are located close to the $K \Gamma K'$ line}
    \label{fig1}
\end{figure*}

\section{Hartree-Fock results: Projected versus fully atomistic model}\label{sec2}

The tight-binding model is defined by the Slater-Koster parametrization of the hopping function of Refs. \cite{moon12,laissandiere10}. We choose a twist angle of $1.05^\circ$, giving a moiré unit cell with 11908 atoms. The relaxation of the atomic lattice is obtained following the model of Ref. \cite{nam17}. Interactions are included via the double-gated Coulomb potential, $V(\rr) = \frac{e^2}{4\pi \epsilon_0 \epsilon_r}\sum_{n=-\infty}^\infty (-1)^n/||\boldsymbol{r} + n\xi \boldsymbol{\hat{z}}||$, regularized at $ \rr=0$ by the Hubbard energy, $U$. We set $\epsilon_r=10$, $U=4$ eV and $\xi=10$ nm. We solve this model in self-consistent Hartree-Fock theory \cite{adhikari24,gonzalez21} at the charge neutrality point, allowing for different symmetry-breaking patterns; then we compare these solutions with those of the flat-band projected theory. For further details on the tight-binding model, the Hartree-Fock method and the projection algorithm, we refer the reader to Appendix \ref{appa} and to Ref. \cite{sanchez25}. 

In Fig. \ref{fig1} we show the band structures of the symmetric normal state (SYM) and several correlated states at charge neutrality: the nematic semimetal (NSM), quantum anomalous Hall (QAH), Kramers intervalley coherent (KIVC), orbital polarized (OP), time-reversal intervalley coherent (TIVC) and valley polarized (VP) states \cite{lian21,bernevig21_2,Bultinck20,Hofmann22,soejima20,wang23,liu21,shi2024moireopticalphononsdancing,wang2024electronphononcouplingtopological,kwan2023electronphononcouplingcompetingkekule}. The normal states of the full tight-binding and projected models are identical by construction, and we find good agreement between the bands of the various symmetry-breaking states. The fully atomistic phases have a slightly larger gap at the $K$ point, from $3$ meV in the VP up to $6$ meV larger in the TIVC state. 

The energies of the different phases are reported in Table \ref{tab1}. The energies of the symmetric states are identical by construction, with minor differences coming from the finer momentum grids used in the projected calculations. In the remaining states, the fully atomistic calculations allow a lower total energy than the flat-band calculations, of about $3$ meV for all phases. 

\begin{table}
    \centering
    \renewcommand{\arraystretch}{1.25} 
    \begin{tabular}{c c c}
        \hline \hline
        \multicolumn{3}{c}{$E - E_\text{SYM}$ (meV)} \\
         & fully atomistic & flat-band projected \\
        \hline
        SYM & $\text{-}$ & \begin{tabular}{@{}c@{}} $-0.01 \ \ (12\times12)$  \\ $-0.03 \ \ (24 \times 24)$ \end{tabular} \\ 
        \hline
        KIVC & $-17.25$ & \begin{tabular}{@{}c@{}} $-14.20 \ \ (12\times12)$ \\ $-14.20 \ \ (24 \times 24)$  \end{tabular}   \\
        \hline
        QAH & $-16.26$ & \begin{tabular}{@{}c@{}} $-12.84 \ \ (12\times12)$ \\ $-12.84 \ \ (24 \times 24)$ \end{tabular}  \\
        \hline
        OP & $-14.82$ & \begin{tabular}{@{}c@{}} $-11.97 \ \ (12\times12)$  \\ $-11.96 \ \ (24 \times 24)$  \end{tabular}  \\
        \hline
        NSM & $-13.17$ & \begin{tabular}{@{}c@{}} $-9.89 \ \ (12\times12)$  \\ $-9.87 \ \ (24 \times 24)$ \end{tabular}  \\
        \hline
        VP & $-12.30$ & \begin{tabular}{@{}c@{}} $-9.11 \ \ (12\times12)$  \\ $-9.17 \ \ (24 \times 24)$ \end{tabular}  \\
        \hline
        TIVC & $-11.15$ & \begin{tabular}{@{}c@{}} $-7.91 \ \ (12\times12)$  \\ $-7.90 \ \ (24 \times 24)$ \end{tabular} \\
        \hline \hline
    \end{tabular}
    \caption{Energies of the normal (SYM) and symmetry-breaking (KIVC, QAH, OP, NSM, VP, TIVC) states with respect to the energy of the fully atomistic (full tight-binding) normal state. The atomistic states were computed on a $6 \times 6$ momentum grid, and we report results for $12 \times 12$ and $24 \times 24$ grids for the flat-band projected model.}
    \label{tab1}
\end{table}

\section{Local order parameters}\label{sec3}

A general wave function at low energies, i.e. supported near the Dirac valleys, can be written as \cite{dossantos12,vafek23}
\begin{align}
    |\psi\rangle =& \sum_\rr \psi(\rr) |\rr \rangle  =\sum_{\eta \sigma \ell} \sum_{\rr \in  \sigma \ell} e^{i \eta \K_\ell \cdot\rr}f_{\sigma \eta \ell}(\rr) |\rr \rangle, 
    \label{wf}
\end{align}
where $\sigma=A,B$ stands for the graphene sublattice, $\eta=+1$$(K),-1$$(K')$ stands for the graphene valley and $\ell=t$(top)$,b$(bottom) stands for the layer. Atoms on sublattice $\sigma$ and layer $\ell$ are denoted by $\rr \in \sigma \ell$, and $\eta \boldsymbol{K}_\ell$ is the $K$ $(\eta=+1)$ or $K'$ $(\eta=-1)$ point of layer $\ell$. The extension to include spin is straightforward. In continuum theories, the envelope functions $f_{\sigma \eta \ell}$ are promoted to functions of continuous space, subject to the normalization condition $\sum_{\eta \sigma \ell} \int d\rr |f_{\eta \sigma \ell}(\rr)|^2 = 1$.

The point group symmetry of MATBG is $C_{6v}$. It includes a two-fold rotation about the $z$ axis, $C_{2z}$, a three-fold rotation about the $z$ axis, $C_{3z}$, and a two-fold rotation about the $x$ axis, $C_{2x}$---acting as a layer-interchanging mirror. Additionally, there is a spinless time reversal symmetry, $\mathcal{T}$, and a $U(1)_v$ symmetry of conservation of the valley charge. They act on the wave functions as follows:
\begin{align}
    C_{2z}&:\ \  \sigma_x \tau_x   &&\rr \to -\rr, \nonumber \\
    C_{3z}&:\ \ e^{-2\pi i /3 \ \sigma_z \tau_z}   && \rr \to R_{-2\pi/3}(\rr), \nonumber \\
    C_{2x}&:\ \ \sigma_x \mu_x  &&\rr \to M_x(\rr), \nonumber \\
    \mathcal{T}&:\ \ \tau_x \mathcal{K}  &&\rr \to \rr, \nonumber \\
    U(1)_v&:\ \ e^{i\varphi  \tau_z}  &&\rr \to \rr,
\end{align}
with $\sigma_i,\tau_i,\mu_i$ denoting the identity ($i=0$) and Pauli ($i=x,y,z$) matrices in sublattice, valley and layer flavor, respectively (sublattice $A$, valley $K$ and the top layer are assigned and eigenvalue of $1$ under their respective Pauli $z$ matrix), $\mathcal{K}$ being the complex conjugation operator and $\varphi$ being an arbitrary angle. $R_{-2\pi/3}$ is the clockwise rotation by angle $2\pi/3$ and $M_x$ is the mirror reflection along the $x$ axis, with the action $\rr = (x,y) \to M_x(\rr) = (x,-y)$.

        

\begin{figure}[t]
    \centering
    \includegraphics[width=.9\linewidth]{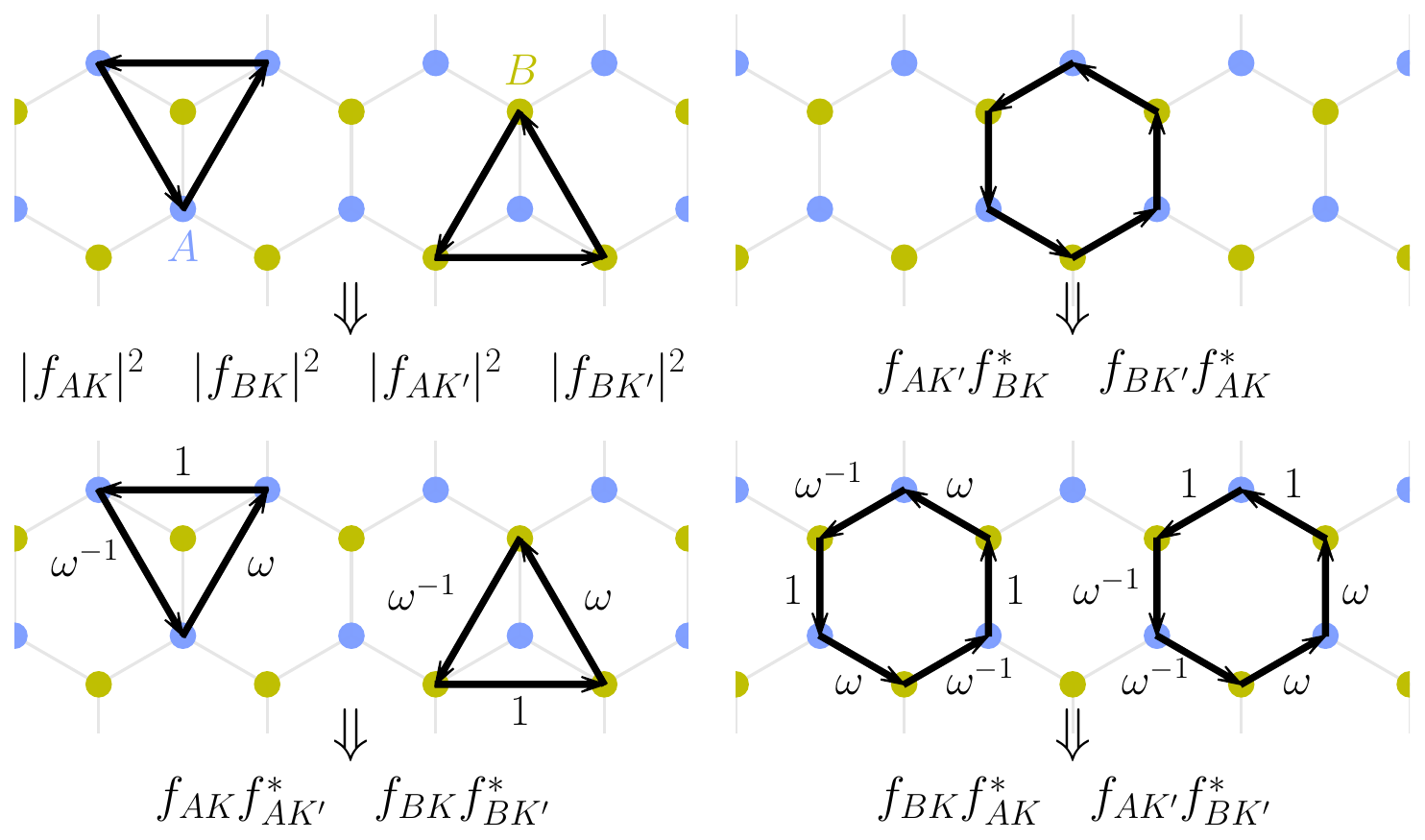}
    \caption{Microscopic loops used to compute the wave function overlaps ($\omega = e^{2\pi i/3}$).}
    \label{loops}
\end{figure}

We compute wave function overlaps of the form
\begin{align}
    f_{\sigma \eta \ell}(\rr) f^*_{\sigma'\eta'\ell}(\rr) 
\end{align}
(note that we consider only intralayer overlaps, with $\ell'=\ell$) by evaluating various types of loops on the lattice, as illustrated in Fig. \ref{loops}, generalizing the original idea of the microscopic valley operator \cite{lopez20,colomes18}. The expressions for the overlaps in terms of the loop quantities are derived in Appendix \ref{appb}.

Summing over the occupied states gives the local order parameters 
\begin{align}
    \rho_{\sigma \eta  \ell,\sigma'\eta'\ell}(\rr) = \sum_{\substack{\text{occupied} \\ \text{states}}} f_{\sigma \eta \ell}(\rr) f_{\sigma'\eta'\ell}^*(\rr).
    \label{rhomatrix}
\end{align}
In this expression the sum over "occupied states" should be understood as the sum over the appropriate indices, like band and momentum, of the occupied states; they are left implicit in order to avoid clutter. We must however restrict the states to be within some energy window around zero energy, such that the expansion in Eq. (\ref{wf}) remains valid. From now on we restrict the sum to run over the occupied states on the flat bands. Results including more bands can be found in Appendix \ref{appc}. 


In Fig. \ref{fig3} we show representative local order parameters of the solutions of the tight-binding model, computed on the top layer. Their distribution is concentrated at the center of the unit cell (the $AA$ region), as expected from the spectral weight of the active orbitals \cite{song22,carr19_2,koshino18}.

In Fig. \ref{fig3}(a) we show the intersublattice, intervalley order $\rho_{BK't,AKt}(\rr)$ of the TIVC and KIVC states. We plot the real part for the TIVC and the imaginary part for the KIVC. In Fig. \ref{fig3}(b) we plot the density on the $A(B)$ sublattice for the OP state, computed as $\sum_\eta \rho_{A(B) \eta t,A(B) \eta t}(\rr)$. The sublattice polarization of this phase is clear. In Fig. \ref{fig3}(c) we show the valley polarization on the $B$ sublattice for QAH and on the $A$ sublattice for VP, computed as $\sum_\eta \eta \rho_{A(B) \eta t,A(B) \eta t}(\rr)$. The sign of the valley polarization is opposite on opposite sublattices for QAH and the same for VP. Finally, in Fig. \ref{fig3}(d) we plot the density on the $A$ sublattice and the intersublattice, intravalley order $\rho_{BKt, AKt}(\rr)$ of the NSM. The breaking of $C_{3z}$ is clear in the density distribution and the phase winding of $\rho_{BKt, AKt}(\rr)$---the phase winds $0$ times around the unit cell versus $-1$ mod $3$ times if $C_{3z}$ were preserved.

\begin{table}[t]
    \centering
    \renewcommand{\arraystretch}{1.25} 
    \begin{tabular}{ c  c  c }
        \hline \hline
         & symmetry & order parameter  \\
        \hline
        KIVC & $C_{3v},\ C_{2z}\mathcal{T}$ & $\langle \sigma_y\tau_x\rangle = 0.838$  \\
        \hline
        QAH & $ C_{6},\ C_{2x}\mathcal{T},\ U(1)_v$ & $\langle  \sigma_z\tau_z\rangle = 0.595$  \\
        \hline
        OP & $C_{3v},\ \mathcal{T}, \  U(1)_v$ & $\langle  \sigma_z \rangle = 0.590$  \\
        \hline
        NSM & $C_{2v},\ \mathcal{T},\ U(1)_v$ & \begin{tabular}{@{}c@{}} $\langle \sigma_x \mu_z \rangle = -0.563$ \\ $\langle \sigma_y\tau_z\mu_z\rangle = 0.331$ \end{tabular}  \\
        \hline
        VP & $C_{3v},\ C _{2z}\mathcal{T},\  U(1)_v$ & $\langle \tau_z  \rangle = 0.886 $ \\
        \hline
        TIVC & $C_{6v},\ \mathcal{T}$ & $\langle \sigma_x \tau_x \rangle = -0.615$ \\
        \hline \hline
    \end{tabular}
    \caption{Symmetry and order parameters of the various symmetry-breaking solutions.}
    \label{tab2}
\end{table}

Integrating over the first unit cell gives the position-independent order parameters \footnote{We note in passing that further summing over the layers corresponds to a partial trace over space and layer of the single-particle density matrix \cite{sanchez24_2},
\begin{align}
    \sum_\ell \rho_{\sigma \eta \ell, \sigma'\eta'\ell} =& \text{Tr}_{\ell \rr}\big( \varrho\big), \\ 
    \varrho_{\sigma\tau\ell,\sigma'\tau'\ell'}(\rr,\rr') =& \sum_{\substack{\text{ occupied} \\ \text{states}}} f_{\sigma \tau \ell}(\rr)  f^*_{\sigma'\eta'\ell'}(\rr'),
\end{align}
with $\text{Tr}_{\ell \rr}(*) \equiv \sum_{\ell \ell'} \delta_{\ell \ell'} \int_{\substack{\text{1º unit } \text{cell}}} d\rr d\rr' \delta(\rr - \rr')(*)$.}   
\begin{align}
    \rho_{\sigma \eta  \ell,\sigma'\eta'\ell} = \int_{\substack{\text{1º unit} \\ \text{cell}}} d\rr \ \rho_{\sigma \eta  \ell,\sigma'\eta'\ell}(\rr). 
\end{align}
Moreover, the different components of $\rho$ can be recast into the following quantities:
\begin{align}
    \langle \sigma_i\tau_j \mu_k \rangle \equiv \frac{1}{2} \text{Tr}\Big(\sigma_i\tau_j\mu_k\rho  \Big),
    \label{orderparams}
\end{align}
with $i,j=0,x,y,z$ and $k=0,z$. With this normalization, the magnitude of $\langle \sigma_i\tau_j \mu_k \rangle$ is bounded by half the number of bands included in the sum of Eq. \ref{rhomatrix}; in particular, by $1$ for the occupied flat bands at the charge neutrality point. In the following, we will omit the identity matrices $\sigma_0,\tau_0,\mu_0$ for brevity.

In Table \ref{tab2} we report the values of the integrated order parameters, as well as the symmetry of the various solutions. The KIVC state breaks $\mathcal{T}$ and $U(1)_v$, while the discrete symmetry consisting of $\mathcal{T}$ followed by a valley rotation with $\varphi=\pi/2$ is preserved \cite{Bultinck20}. $C_{2z}$ is also broken, but the product $C_{2z}\mathcal{T}$ is preserved. QAH breaks the mirror symmetries and time-reversal while preserving $C_{2x}\mathcal{T}$. OP and NSM break $C_{2z}$ and $C_{3z}$, respectively. VP breaks $C_{2z}$ and $\mathcal{T}$ while preserving $C_{2z}\mathcal{T}$. Finally, the TIVC state breaks $U(1)_v$ but preserves all point symmetries and the time-reversal symmetry.

The order parameters are constrained by the symmetry and reflect defining physical features of each state. The intervalley coherent states display wave functions with superpositions of opposite valleys and sublattices, with nonzero $\langle \sigma_y \tau_x \rangle$ and $\langle \sigma_x \tau_x \rangle$ in KIVC and TIVC, respectively. The OP state shows sublattice polarization, $\langle \sigma_z \rangle$, and the VP state shows valley polarization, $\langle \tau_z \rangle$. The QAH state has the order parameter $\langle \sigma_z \tau_z \rangle$, with opposite valley polarizations on opposite sublattices. This is related to a nonzero Chern number by the properties of the flat-band wave functions \cite{Bultinck20,sanchez25} (the Chern number per spin projection is $2$). Finally, NSM displays the $C_{3z}$-breaking quantities $ \langle \sigma_x \mu_z \rangle$ and $\langle \sigma_y \tau_z \mu_z \rangle$. The mirror $C_{3z}^{} C_{2x} C_{3z}^{-1}$ further sets the phase of $\langle \sigma_x\mu_z \rangle + i \langle \sigma_y\tau_z\mu_z \rangle$ to $5\pi/6 \ \text{mod} \ \pi$.

Moreover, $\langle \sigma_0 \tau_0 \mu_0 \rangle$ counts half the number of occupied bands (we now write the identity matrices explicitly), and we consistently get $\langle \sigma_0 \tau_0 \mu_0 \rangle = 0.997$ in all phases. All remaining order parameters are negligible, except only in the NSM state where we find the subdominant values $\langle \sigma_x \rangle = -0.0043$ and $\langle \sigma_y \tau_x \rangle = -0.0071$.


\section{DISCUSSION}\label{sec4}

With the aim of benchmarking our projection algorithm \cite{sanchez25}, we performed self-consistent Hartree-Fock calculations on an interacting tight-binding model of magic-angle twisted bilayer graphene (MATBG). The results for the flat-band projected and fully atomistic models show good agreement, thereby validating the projection scheme and pointing to the reliability of the flat-band projection \cite{ledwith25}. In fact, the splitting of the flat bands in the symmetry-breaking phases (to $\pm \ 15 \text{-} 20$ meV) is less than half the energies of the band edges of the remote bands in the normal state---about $+60$ meV and $-50$ meV for the conduction and valence bands, respectively. This separation of scales indicates that the remote bands are effectively frozen. By contrast, in the noninteracting state the remote bands appear at about $\pm 25$ meV, i.e. at energies comparable to the flat-band splitting \cite{wang2024electronphononcouplingtopological,song22,shi2024moireopticalphononsdancing}. 

We emphasize that the many-body projection correctly captures the interacting effects of integrated-out bands, including the increase of the flat-band bandwidth of the normal state. A naive truncation to the flat manifold, which neglects the remote-band contributions, will fail to reproduce these features and the results of the fully atomistic theory.     

\begin{figure}[t]
    \centering
    \includegraphics[width=0.65\linewidth]{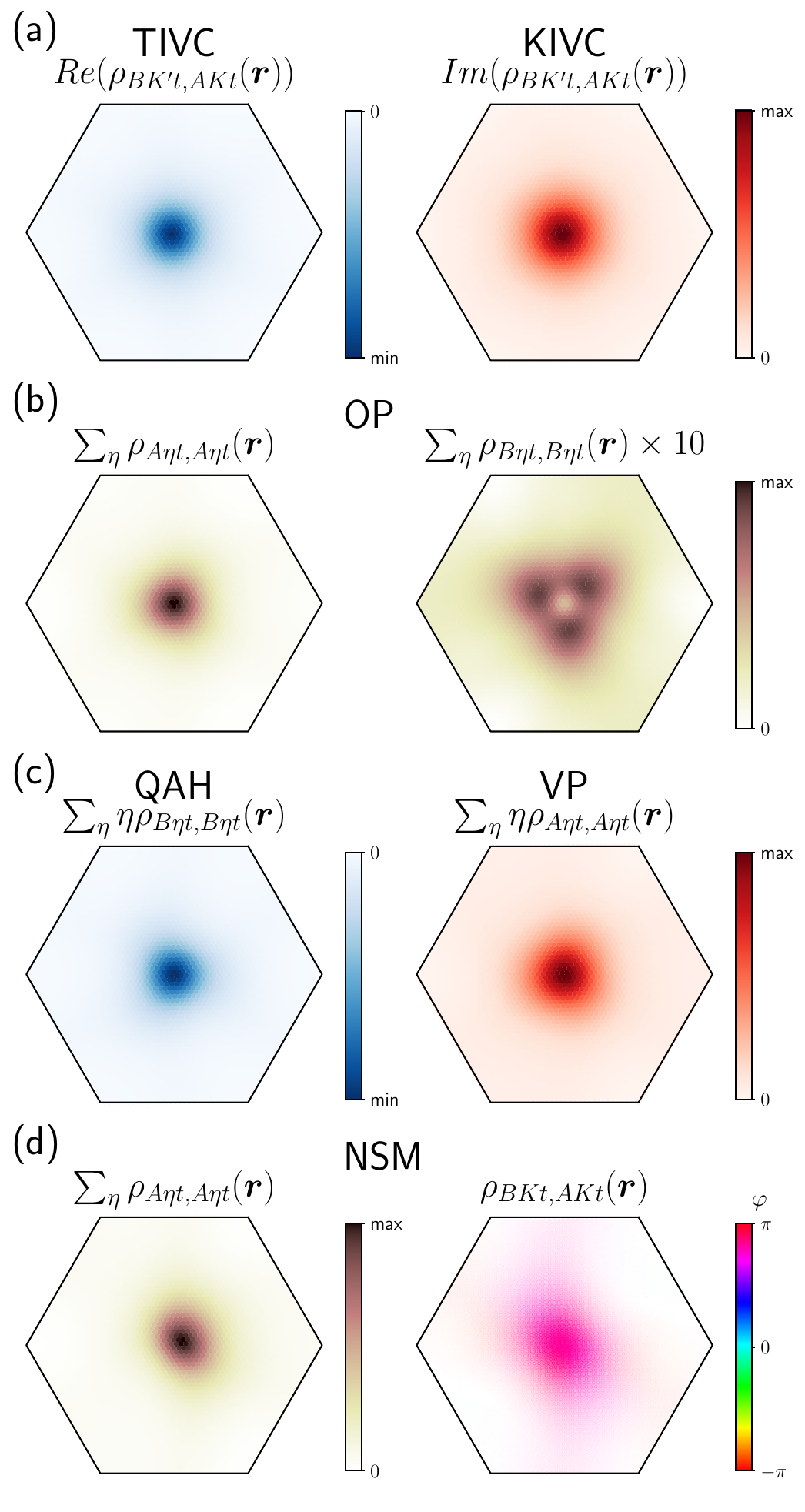}
    \caption{Local order parameters. (a) Intrasublattice, intervalley coherence of TIVC (real part) and KIVC (imaginary part). (b) Density of the OP state on the two sublattices; the density on the $B$ sublattice is multiplied by $10$ for better visualization. (c) Valley polarization on sublattice $B$ for QAH and sublattice $A$ for VP. (d) NSM state. Left: density on sublattice $A$. Right: intersublattice, intravalley order parameter. the transparency and color indicate the magnitude and the phase of the complex number, respectively. All quantities are shown for the top layer.}
    \label{fig3}
\end{figure}

Several remarks are in order regarding the modeling of the electronic interactions. The interaction strength is governed by the effective dielectric constant $\epsilon_r$ and the on-site Hubbard $U$, which are free parameters in our setup. A suitable choice of $\epsilon_r$ should take into account the effects of the screening from both the dielectric substrate and the internal electronic screening. Random phase approximation (RPA) calculations of the dielectric function \cite{Pizarro19,vanhala20} yield $\epsilon_r(q) \sim 10$ for momenta $q \gtrsim 1 \ \text{nm}^{-1}$, or equivalently for distances smaller than about $1 \ \text{nm}$. This is consistent with the value for $8$ Dirac fermions ($2$ spins$\times \ 2$ valleys $\times \ 2$ layers) at charge neutrality within a hexagonal boron nitride (hBN) environment,
\begin{align}
    \epsilon_r = \epsilon_{hBN} + 8 \bigg(\frac{e^2}{32\epsilon_0 v_F}\bigg) \sim 12.6,  
    \label{eqepsilon}
\end{align}
with $v_F$ being the Fermi velocity of graphene, and setting $\epsilon_{hBN} = 4$. For larger distances, virtual transitions between the flat and dispersive bands increase the dielectric function up to $\epsilon_r(q) \sim 20$ at $q \approx 0.25 \ \text{nm}^{-1}$ near the magic angle \cite{vanhala20}. For even larger distances the polarization depends on the details of the flat bands, ranging from a Thomas-Fermi screened behavior if the system is metallic to the graphene-like form of Eq. (\ref{eqepsilon}), with $v_F$ replaced by the Fermi velocity of the flat bands, when the twist angle is not near the magic angle \cite{Pizarro19}. In this work, we fix $\epsilon_r=10$ as in Refs. \cite{wang23,kwan21,kwan2023electronphononcouplingcompetingkekule,wang2024electronphononcouplingtopological,shi2024moireopticalphononsdancing}; similar values of $\epsilon_r = 12\text{-}16$ have been used in order to match experimental \cite{wang24,Choi2021} and numerical \cite{calderón2025cascadestransportopticalconductivity, calderon20} data. Precise treatment of the dielectric function, like in GW theory \cite{zhu24,peng2025manybodyperturbationtheorymoire,Romanova2022,guandalini24}, would represent critical improvements of our results, particularly in the construction of the normal state and projected models.

The gate distance and number of gates of the experimental device also modify the electronic interactions \cite{Lau25, calderon20}, and experiments have shown that changing the device geometry can tune the superconducting and insulating regimes \cite{Stepanov2020,Saito2020}. We adopt a double-gated Coulomb potential with a gate-to-gate distance of $\xi=10$ nm, alongside other theoretical works \cite{song22,rai24,calugaru2025obtainingspectralfunctionmoire,adhikari24,bernevig21,lian21,wang24_2}. 

Following Ref. \cite{Lau25}, the Coulomb integral between two flat-band orbitals is estimated as
\begin{align}
    U_1 \sim \frac{1792 \text{ meV}}{\epsilon_r \ \lambda(\text{nm})}  F\bigg(\frac{\xi}{2\lambda}\bigg),
\end{align}
where $\lambda = 2.9$ nm characterizes the orbital localization and $F(x)$ is a function characterizing the screening by the gates. This gives $U_1 \sim 42$ meV, comparable to the gaps at the $K$ points of the states in Fig. \ref{fig1}, ranging from $30.8$ meV in the flat-band projected OP state to $38.8$ meV in the atomistic TIVC state. This scale is also comparable to the observed separation of the local density of states peaks in tunneling experiments of $25\text{-}40$ meV  \cite{Wong2020,Xie2019,Jiang2019,Choi2019}, and to the gap at $K$ of $\sim 35$ meV in the quantum twisting microscopy experiment of Ref. \cite{xiao2025interactingenergybandsmagic}. We must stress, however, that in those experiments the gate distance is usually hundreds of nm, leading to $F(\xi/2\lambda) \approx 1$ and $U_1 \sim 62$ meV. Dynamical screening effects further renormalize the gap down, by about $40 \%$ in Refs. \cite{crippa2025dynamicalcorrelationeffectstwisted,rai24, calugaru2025obtainingspectralfunctionmoire}, recovering the scale of $25\text{-}40$ meV with our choice of $\epsilon_r=10$.

With respect to the Hubbard energy, we fix $U=4$ eV, within the range $3\text{-}9$ eV \cite{wehling11,schuler13}. In MATBG the on-site interaction is weak with respect to the long-range interaction \cite{Guinea2018} and enters as a degeneracy-lifting perturbation. The large variety of nearly degenerate ground-state candidates in MATBG turns the effects from the on-site interactions (and especially the electron-phonon coupling \cite{shi2024moireopticalphononsdancing,kwan2023electronphononcouplingcompetingkekule,Chen2024,angeli19,zhao25}) relevant in selecting the ground state  \cite{wang2024electronphononcouplingtopological}, which can be superconducting \cite{wang24_2}. Such effects are beyond the scope of this work.

On a different note, we employed novel algorithms for computing general wave function overlaps in valley-sublattice space, first introduced in Ref. \cite{sanchez24_2}. With them we probe the symmetry breaking of MATBG locally in real space, complementing the information coming from the one-particle density matrix in momentum space \cite{sanchez25, Bultinck20, liu21, kwan21, kwan2023electronphononcouplingcompetingkekule, Hofmann22, soejima20, wagner22, wang23}. 

Integrating the local wave function overlaps over the unit cell provides a set of order parameters for the various solutions. We find that the magnitude of the integrated order parameters computed on the flat bands is comparable to the theoretical maximum of $1$. In fact, at the chosen twist angle of $1.05^\circ$, due to the increased bandwidth of the parent normal state, the electrons near the $\Gamma$ point do not break the symmetry, as noted in \cite{sanchez25}. We expect then that the order parameters will attain values close to saturation at a lower angle where the flat bands reappear \cite{sánchez2025fermivelocitymagicangle}. Furthermore, when we include the remote bands their numerical values are not greatly modified (see Appendix \ref{appc}), revealing that the symmetry breaking is strongest on the flat bands. 

The local order parameters can be computed directly in atomistic models, without the need of constructing the Chern basis of momentum space descriptions \cite{Bultinck20, liu21, kwan21, kwan2023electronphononcouplingcompetingkekule, Hofmann22, soejima20, wagner22, wang23}; the remote bands can also be naturally included in this setup. Overall, our algorithms for obtaining wave function overlaps extend the analytical toolkit for the study of MATBG and other honeycomb systems. A natural extension of our work would be to derive expressions for the interlayer overlaps.



\section{Acknowledgments}
The work was supported by Grants No.PID2023-146461NB-I00, funded by Ministerio de Ciencia, Innovaci\'on y Universidades, and Grant No. PRE2021-097070, funded by Ministerio de Ciencia, Innovaci\'on y Universidades through Agencia Estatal de Investigaci\'on, as well as by the CSIC Research Platform on Quantum Technologies PTI-001 and the Severo Ochoa Centres of Excellence program through Grant CEX2024-001445-S. The access to the computational resources of CESGA (Centro de Supercomputaci\'on de Galicia) is also gratefully acknowledged.

\clearpage

\bibliography{biblio}

@PREAMBLE{
"\UseRawInputEncoding"
}

@article{gonzalez19,
  title = {Kohn-Luttinger Superconductivity in Twisted Bilayer Graphene},
  author = {Gonz\'alez, J. and Stauber, T.},
  journal = {Phys. Rev. Lett.},
  volume = {122},
  issue = {2},
  pages = {026801},
  numpages = {6},
  year = {2019},
  month = {Jan},
  publisher = {American Physical Society},
  doi = {10.1103/PhysRevLett.122.026801},
  url = {https://link.aps.org/doi/10.1103/PhysRevLett.122.026801}
}

@article{Cao2018,
  title = {Unconventional superconductivity in magic-angle graphene superlattices},
  volume = {556},
  ISSN = {1476-4687},
  url = {http://dx.doi.org/10.1038/nature26160},
  DOI = {10.1038/nature26160},
  number = {7699},
  journal = {Nature},
  publisher = {Springer Science and Business Media LLC},
  author = {Cao,  Yuan and Fatemi,  Valla and Fang,  Shiang and Watanabe,  Kenji and Taniguchi,  Takashi and Kaxiras,  Efthimios and Jarillo-Herrero,  Pablo},
  year = {2018},
  month = mar,
  pages = {43–50}
}

@article{Cao2018_2,
  title = {Correlated insulator behaviour at half-filling in magic-angle graphene superlattices},
  volume = {556},
  ISSN = {1476-4687},
  url = {http://dx.doi.org/10.1038/nature26154},
  DOI = {10.1038/nature26154},
  number = {7699},
  journal = {Nature},
  publisher = {Springer Science and Business Media LLC},
  author = {Cao,  Yuan and Fatemi,  Valla and Demir,  Ahmet and Fang,  Shiang and Tomarken,  Spencer L. and Luo,  Jason Y. and Sanchez-Yamagishi,  Javier D. and Watanabe,  Kenji and Taniguchi,  Takashi and Kaxiras,  Efthimios and Ashoori,  Ray C. and Jarillo-Herrero,  Pablo},
  year = {2018},
  month = mar,
  pages = {80–84}
}

@article{Mak2022,
  title = {Semiconductor moir\'e materials},
  volume = {17},
  ISSN = {1748-3395},
  url = {http://dx.doi.org/10.1038/s41565-022-01165-6},
  DOI = {10.1038/s41565-022-01165-6},
  number = {7},
  journal = {Nature Nanotechnology},
  publisher = {Springer Science and Business Media LLC},
  author = {Mak,  Kin Fai and Shan,  Jie},
  year = {2022},
  month = jul,
  pages = {686–695}
}

@article{Nuckolls2024,
  title = {A microscopic perspective on moir\'e materials},
  volume = {9},
  ISSN = {2058-8437},
  url = {http://dx.doi.org/10.1038/s41578-024-00682-1},
  DOI = {10.1038/s41578-024-00682-1},
  number = {7},
  journal = {Nature Reviews Materials},
  publisher = {Springer Science and Business Media LLC},
  author = {Nuckolls,  Kevin P. and Yazdani,  Ali},
  year = {2024},
  month = may,
  pages = {460–480}
}

@article{Kennes2021,
  title = {Moiré heterostructures as a condensed-matter quantum simulator},
  volume = {17},
  ISSN = {1745-2481},
  url = {http://dx.doi.org/10.1038/s41567-020-01154-3},
  DOI = {10.1038/s41567-020-01154-3},
  number = {2},
  journal = {Nature Physics},
  publisher = {Springer Science and Business Media LLC},
  author = {Kennes,  Dante M. and Claassen,  Martin and Xian,  Lede and Georges,  Antoine and Millis,  Andrew J. and Hone,  James and Dean,  Cory R. and Basov,  D. N. and Pasupathy,  Abhay N. and Rubio,  Angel},
  year = {2021},
  month = feb,
  pages = {155–163}
}

@article{wang24,
  title = {Theory of Correlated Chern Insulators in Twisted Bilayer Graphene},
  author = {Wang, Xiaoyu and Vafek, Oskar},
  journal = {Phys. Rev. X},
  volume = {14},
  issue = {2},
  pages = {021042},
  numpages = {15},
  year = {2024},
  month = {Jun},
  publisher = {American Physical Society},
  doi = {10.1103/PhysRevX.14.021042},
  url = {https://link.aps.org/doi/10.1103/PhysRevX.14.021042}
}

@article{song22,
  title = {Magic-Angle Twisted Bilayer Graphene as a Topological Heavy Fermion Problem},
  author = {Song, Zhi-Da and Bernevig, B. Andrei},
  journal = {Phys. Rev. Lett.},
  volume = {129},
  issue = {4},
  pages = {047601},
  numpages = {10},
  year = {2022},
  month = {Jul},
  publisher = {American Physical Society},
  doi = {10.1103/PhysRevLett.129.047601},
  url = {https://link.aps.org/doi/10.1103/PhysRevLett.129.047601}
}

@article{ledwith25,
  title = {Nonlocal Moments and Mott Semimetal in the Chern Bands of Twisted Bilayer Graphene},
  author = {Ledwith, Patrick J. and Dong, Junkai and Vishwanath, Ashvin and Khalaf, Eslam},
  journal = {Phys. Rev. X},
  volume = {15},
  issue = {2},
  pages = {021087},
  numpages = {40},
  year = {2025},
  month = {Jun},
  publisher = {American Physical Society},
  doi = {10.1103/PhysRevX.15.021087},
  url = {https://link.aps.org/doi/10.1103/PhysRevX.15.021087}
}

@article{wehling11,
  title = {Strength of Effective Coulomb Interactions in Graphene and Graphite},
  author = {Wehling, T. O. and \ifmmode \mbox{\c{S}}\else \c{S}\fi{}a\ifmmode \mbox{\c{s}}\else \c{s}\fi{}\ifmmode \imath \else \i \fi{}o\ifmmode \breve{g}\else \u{g}\fi{}lu, E. and Friedrich, C. and Lichtenstein, A. I. and Katsnelson, M. I. and Bl\"ugel, S.},
  journal = {Phys. Rev. Lett.},
  volume = {106},
  issue = {23},
  pages = {236805},
  numpages = {4},
  year = {2011},
  month = {Jun},
  publisher = {American Physical Society},
  doi = {10.1103/PhysRevLett.106.236805},
  url = {https://link.aps.org/doi/10.1103/PhysRevLett.106.236805}
}

@article{schuler13,
  title = {Optimal Hubbard Models for Materials with Nonlocal Coulomb Interactions: Graphene, Silicene, and Benzene},
  author = {Sch\"uler, M. and R\"osner, M. and Wehling, T. O. and Lichtenstein, A. I. and Katsnelson, M. I.},
  journal = {Phys. Rev. Lett.},
  volume = {111},
  issue = {3},
  pages = {036601},
  numpages = {5},
  year = {2013},
  month = {Jul},
  publisher = {American Physical Society},
  doi = {10.1103/PhysRevLett.111.036601},
  url = {https://link.aps.org/doi/10.1103/PhysRevLett.111.036601}
}

@article{guandalini24,
  title = {Efficient $GW$ calculations via interpolation of the screened interaction in momentum and frequency space: The case of graphene},
  author = {Guandalini, Alberto and Leon, Dario A. and D'Amico, Pino and Cardoso, Claudia and Ferretti, Andrea and Rontani, Massimo and Varsano, Daniele},
  journal = {Phys. Rev. B},
  volume = {109},
  issue = {7},
  pages = {075120},
  numpages = {15},
  year = {2024},
  month = {Feb},
  publisher = {American Physical Society},
  doi = {10.1103/PhysRevB.109.075120},
  url = {https://link.aps.org/doi/10.1103/PhysRevB.109.075120}
}

@misc{calderón2025cascadestransportopticalconductivity,
      title={Cascades in transport and optical conductivity of Twisted Bilayer Graphene}, 
      author={M. J. Calderón and A. Camjayi and A. Datta and E. Bascones},
      year={2025},
      eprint={2412.20855},
      archivePrefix={arXiv},
      primaryClass={cond-mat.str-el},
      url={https://arxiv.org/abs/2412.20855}, 
}

@article{Guinea2018,
  title = {Electrostatic effects,  band distortions,  and superconductivity in twisted graphene bilayers},
  volume = {115},
  ISSN = {1091-6490},
  url = {http://dx.doi.org/10.1073/pnas.1810947115},
  DOI = {10.1073/pnas.1810947115},
  number = {52},
  journal = {Proceedings of the National Academy of Sciences},
  publisher = {Proceedings of the National Academy of Sciences},
  author = {Guinea,  Francisco and Walet,  Niels R.},
  year = {2018},
  month = dec,
  pages = {13174–13179}
}

@article{huang25,
  title = {Perturbative renormalization group approach to magic-angle twisted bilayer graphene using topological heavy fermion model},
  author = {Huang, Yi and Chou, Yang-Zhi and Das Sarma, Sankar},
  journal = {Phys. Rev. B},
  volume = {112},
  issue = {24},
  pages = {245132},
  numpages = {20},
  year = {2025},
  month = {Dec},
  publisher = {American Physical Society},
  doi = {10.1103/2px8-h7c2},
  url = {https://link.aps.org/doi/10.1103/2px8-h7c2}
}

@article{sanchez25,
  title = {Nonflat bands and chiral symmetry in magic-angle twisted bilayer graphene},
  author = {S\'anchez S\'anchez, Miguel and Gonz\'alez, Jos\'e and Stauber, Tobias},
  journal = {Phys. Rev. B},
  volume = {111},
  issue = {20},
  pages = {205133},
  numpages = {19},
  year = {2025},
  month = {May},
  publisher = {American Physical Society},
  doi = {10.1103/PhysRevB.111.205133},
  url = {https://link.aps.org/doi/10.1103/PhysRevB.111.205133}
}

@article{Bultinck20,
  title = {Ground State and Hidden Symmetry of Magic-Angle Graphene at Even Integer Filling},
  author = {Bultinck, Nick and Khalaf, Eslam and Liu, Shang and Chatterjee, Shubhayu and Vishwanath, Ashvin and Zaletel, Michael P.},
  journal = {Phys. Rev. X},
  volume = {10},
  issue = {3},
  pages = {031034},
  numpages = {13},
  year = {2020},
  month = {Aug},
  publisher = {American Physical Society},
  doi = {10.1103/PhysRevX.10.031034},
  url = {https://link.aps.org/doi/10.1103/PhysRevX.10.031034}
}

@article{wagner22,
  title = {Global Phase Diagram of the Normal State of Twisted Bilayer Graphene},
  author = {Wagner, Glenn and Kwan, Yves H. and Bultinck, Nick and Simon, Steven H. and Parameswaran, S. A.},
  journal = {Phys. Rev. Lett.},
  volume = {128},
  issue = {15},
  pages = {156401},
  numpages = {7},
  year = {2022},
  month = {Apr},
  publisher = {American Physical Society},
  doi = {10.1103/PhysRevLett.128.156401},
  url = {https://link.aps.org/doi/10.1103/PhysRevLett.128.156401}
}

@article{vafek20,
  title = {Renormalization Group Study of Hidden Symmetry in Twisted Bilayer Graphene with Coulomb Interactions},
  author = {Vafek, Oskar and Kang, Jian},
  journal = {Phys. Rev. Lett.},
  volume = {125},
  issue = {25},
  pages = {257602},
  numpages = {7},
  year = {2020},
  month = {Dec},
  publisher = {American Physical Society},
  doi = {10.1103/PhysRevLett.125.257602},
  url = {https://link.aps.org/doi/10.1103/PhysRevLett.125.257602}
}

@article{vafek23,
  title = {Continuum effective Hamiltonian for graphene bilayers for an arbitrary smooth lattice deformation from microscopic theories},
  author = {Vafek, Oskar and Kang, Jian},
  journal = {Phys. Rev. B},
  volume = {107},
  issue = {7},
  pages = {075123},
  numpages = {11},
  year = {2023},
  month = {Feb},
  publisher = {American Physical Society},
  doi = {10.1103/PhysRevB.107.075123},
  url = {https://link.aps.org/doi/10.1103/PhysRevB.107.075123}
}

@article{kwan21,
  title = {Kekul\'e Spiral Order at All Nonzero Integer Fillings in Twisted Bilayer Graphene},
  author = {Kwan, Y. H. and Wagner, G. and Soejima, T. and Zaletel, M. P. and Simon, S. H. and Parameswaran, S. A. and Bultinck, N.},
  journal = {Phys. Rev. X},
  volume = {11},
  issue = {4},
  pages = {041063},
  numpages = {23},
  year = {2021},
  month = {Dec},
  publisher = {American Physical Society},
  doi = {10.1103/PhysRevX.11.041063},
  url = {https://link.aps.org/doi/10.1103/PhysRevX.11.041063}
}

@article{Tian2023,
  title = {Evidence for Dirac flat band superconductivity enabled by quantum geometry},
  volume = {614},
  ISSN = {1476-4687},
  url = {http://dx.doi.org/10.1038/s41586-022-05576-2},
  DOI = {10.1038/s41586-022-05576-2},
  number = {7948},
  journal = {Nature},
  publisher = {Springer Science and Business Media LLC},
  author = {Tian,  Haidong and Gao,  Xueshi and Zhang,  Yuxin and Che,  Shi and Xu,  Tianyi and Cheung,  Patrick and Watanabe,  Kenji and Taniguchi,  Takashi and Randeria,  Mohit and Zhang,  Fan and Lau,  Chun Ning and Bockrath,  Marc W.},
  year = {2023},
  month = feb,
  pages = {440–444}
}

@article{wang23,
  title = {Ground-state order in magic-angle graphene at filling $\ensuremath{\nu}=\ensuremath{-}3$: A full-scale density matrix renormalization group study},
  author = {Wang, Tianle and Parker, Daniel E. and Soejima, Tomohiro and Hauschild, Johannes and Anand, Sajant and Bultinck, Nick and Zaletel, Michael P.},
  journal = {Phys. Rev. B},
  volume = {108},
  issue = {23},
  pages = {235128},
  numpages = {8},
  year = {2023},
  month = {Dec},
  publisher = {American Physical Society},
  doi = {10.1103/PhysRevB.108.235128},
  url = {https://link.aps.org/doi/10.1103/PhysRevB.108.235128}
}

@article{shi2024moireopticalphononsdancing,
  title = {Moir\'e optical phonons coupled to heavy electrons in magic-angle twisted bilayer graphene},
  author = {Shi, Hao and Miao, Wangqian and Dai, Xi},
  journal = {Phys. Rev. B},
  volume = {111},
  issue = {15},
  pages = {155126},
  numpages = {33},
  year = {2025},
  month = {Apr},
  publisher = {American Physical Society},
  doi = {10.1103/PhysRevB.111.155126},
  url = {https://link.aps.org/doi/10.1103/PhysRevB.111.155126}
}

@article{Zondiner2020,
  title = {Cascade of phase transitions and Dirac revivals in magic-angle graphene},
  volume = {582},
  ISSN = {1476-4687},
  url = {http://dx.doi.org/10.1038/s41586-020-2373-y},
  DOI = {10.1038/s41586-020-2373-y},
  number = {7811},
  journal = {Nature},
  publisher = {Springer Science and Business Media LLC},
  author = {Zondiner,  U. and Rozen,  A. and Rodan-Legrain,  D. and Cao,  Y. and Queiroz,  R. and Taniguchi,  T. and Watanabe,  K. and Oreg,  Y. and von Oppen,  F. and Stern,  Ady and Berg,  E. and Jarillo-Herrero,  P. and Ilani,  S.},
  year = {2020},
  month = jun,
  pages = {203–208}
}

@article{cao2020,
  title = {Strange Metal in Magic-Angle Graphene with near Planckian Dissipation},
  author = {Cao, Yuan and Chowdhury, Debanjan and Rodan-Legrain, Daniel and Rubies-Bigorda, Oriol and Watanabe, Kenji and Taniguchi, Takashi and Senthil, T. and Jarillo-Herrero, Pablo},
  journal = {Phys. Rev. Lett.},
  volume = {124},
  issue = {7},
  pages = {076801},
  numpages = {7},
  year = {2020},
  month = {Feb},
  publisher = {American Physical Society},
  doi = {10.1103/PhysRevLett.124.076801},
  url = {https://link.aps.org/doi/10.1103/PhysRevLett.124.076801}
}

@article{Polshyn2019,
  title = {Large linear-in-temperature resistivity in twisted bilayer graphene},
  volume = {15},
  ISSN = {1745-2481},
  url = {http://dx.doi.org/10.1038/s41567-019-0596-3},
  DOI = {10.1038/s41567-019-0596-3},
  number = {10},
  journal = {Nature Physics},
  publisher = {Springer Science and Business Media LLC},
  author = {Polshyn,  Hryhoriy and Yankowitz,  Matthew and Chen,  Shaowen and Zhang,  Yuxuan and Watanabe,  K. and Taniguchi,  T. and Dean,  Cory R. and Young,  Andrea F.},
  year = {2019},
  month = aug,
  pages = {1011–1016}
}

@article{Jaoui2022,
  title = {Quantum critical behaviour in magic-angle twisted bilayer graphene},
  volume = {18},
  ISSN = {1745-2481},
  url = {http://dx.doi.org/10.1038/s41567-022-01556-5},
  DOI = {10.1038/s41567-022-01556-5},
  number = {6},
  journal = {Nature Physics},
  publisher = {Springer Science and Business Media LLC},
  author = {Jaoui,  Alexandre and Das,  Ipsita and Di Battista,  Giorgio and Díez-Mérida,  Jaime and Lu,  Xiaobo and Watanabe,  Kenji and Taniguchi,  Takashi and Ishizuka,  Hiroaki and Levitov,  Leonid and Efetov,  Dmitri K.},
  year = {2022},
  month = apr,
  pages = {633–638}
}

@article{Choi2019,
  title = {Electronic correlations in twisted bilayer graphene near the magic angle},
  volume = {15},
  ISSN = {1745-2481},
  url = {http://dx.doi.org/10.1038/s41567-019-0606-5},
  DOI = {10.1038/s41567-019-0606-5},
  number = {11},
  journal = {Nature Physics},
  publisher = {Springer Science and Business Media LLC},
  author = {Choi,  Youngjoon and Kemmer,  Jeannette and Peng,  Yang and Thomson,  Alex and Arora,  Harpreet and Polski,  Robert and Zhang,  Yiran and Ren,  Hechen and Alicea,  Jason and Refael,  Gil and von Oppen,  Felix and Watanabe,  Kenji and Taniguchi,  Takashi and Nadj-Perge,  Stevan},
  year = {2019},
  month = aug,
  pages = {1174–1180}
}

@article{tomarken19,
  title = {Electronic Compressibility of Magic-Angle Graphene Superlattices},
  author = {Tomarken, S. L. and Cao, Y. and Demir, A. and Watanabe, K. and Taniguchi, T. and Jarillo-Herrero, P. and Ashoori, R. C.},
  journal = {Phys. Rev. Lett.},
  volume = {123},
  issue = {4},
  pages = {046601},
  numpages = {6},
  year = {2019},
  month = {Jul},
  publisher = {American Physical Society},
  doi = {10.1103/PhysRevLett.123.046601},
  url = {https://link.aps.org/doi/10.1103/PhysRevLett.123.046601}
}

@article{liu21,
  title = {Nematic topological semimetal and insulator in magic-angle bilayer graphene at charge neutrality},
  author = {Liu, Shang and Khalaf, Eslam and Lee, Jong Yeon and Vishwanath, Ashvin},
  journal = {Phys. Rev. Res.},
  volume = {3},
  issue = {1},
  pages = {013033},
  numpages = {12},
  year = {2021},
  month = {Jan},
  publisher = {American Physical Society},
  doi = {10.1103/PhysRevResearch.3.013033},
  url = {https://link.aps.org/doi/10.1103/PhysRevResearch.3.013033}
}

@article{kwan2023electronphononcouplingcompetingkekule,
  title = {Electron-phonon coupling and competing Kekul\'e orders in twisted bilayer graphene},
  author = {Kwan, Yves H. and Wagner, Glenn and Bultinck, Nick and Simon, Steven H. and Berg, Erez and Parameswaran, S. A.},
  journal = {Phys. Rev. B},
  volume = {110},
  issue = {8},
  pages = {085160},
  numpages = {9},
  year = {2024},
  month = {Aug},
  publisher = {American Physical Society},
  doi = {10.1103/PhysRevB.110.085160},
  url = {https://link.aps.org/doi/10.1103/PhysRevB.110.085160}
}

@article{wang2024electronphononcouplingtopological,
  title = {Electron-phonon coupling in the topological heavy fermion model of twisted bilayer graphene},
  author = {Wang, Yi-Jie and Zhou, Geng-Dong and Lian, Biao and Song, Zhi-Da},
  journal = {Phys. Rev. B},
  volume = {111},
  issue = {3},
  pages = {035110},
  numpages = {47},
  year = {2025},
  month = {Jan},
  publisher = {American Physical Society},
  doi = {10.1103/PhysRevB.111.035110},
  url = {https://link.aps.org/doi/10.1103/PhysRevB.111.035110}
}

@article{Lau25,
  title = {Topological Mixed Valence Model for Twisted Bilayer Graphene},
  author = {Lau, Liam L. H. and Coleman, Piers},
  journal = {Phys. Rev. X},
  volume = {15},
  issue = {2},
  pages = {021028},
  numpages = {29},
  year = {2025},
  month = {Apr},
  publisher = {American Physical Society},
  doi = {10.1103/PhysRevX.15.021028},
  url = {https://link.aps.org/doi/10.1103/PhysRevX.15.021028}
}

@misc{calugaru2025obtainingspectralfunctionmoire,
      title={Obtaining the Spectral Function of Moir\'e Graphene Heavy-Fermions Using Iterative Perturbation Theory}, 
      author={Dumitru Călugăru and Haoyu Hu and Lorenzo Crippa and Gautam Rai and Nicolas Regnault and Tim O. Wehling and Roser Valentí and Giorgio Sangiovanni and B. Andrei Bernevig},
      year={2025},
      eprint={2509.18256},
      archivePrefix={arXiv},
      primaryClass={cond-mat.str-el},
      url={https://arxiv.org/abs/2509.18256}, 
}

@misc{crippa2025dynamicalcorrelationeffectstwisted,
      title={Dynamical correlation effects in twisted bilayer graphene under strain and lattice relaxation}, 
      author={Lorenzo Crippa and Gautam Rai and Dumitru Călugăru and Haoyu Hu and Jonah Herzog-Arbeitman and B. Andrei Bernevig and Roser Valentí and Giorgio Sangiovanni and Tim Wehling},
      year={2025},
      eprint={2509.19436},
      archivePrefix={arXiv},
      primaryClass={cond-mat.str-el},
      url={https://arxiv.org/abs/2509.19436}, 
}

@article{Choi2021,
  title = {Interaction-driven band flattening and correlated phases in twisted bilayer graphene},
  volume = {17},
  ISSN = {1745-2481},
  url = {http://dx.doi.org/10.1038/s41567-021-01359-0},
  DOI = {10.1038/s41567-021-01359-0},
  number = {12},
  journal = {Nature Physics},
  publisher = {Springer Science and Business Media LLC},
  author = {Choi,  Youngjoon and Kim,  Hyunjin and Lewandowski,  Cyprian and Peng,  Yang and Thomson,  Alex and Polski,  Robert and Zhang,  Yiran and Watanabe,  Kenji and Taniguchi,  Takashi and Alicea,  Jason and Nadj-Perge,  Stevan},
  year = {2021},
  month = nov,
  pages = {1375–1381}
}

@article{Hofmann22,
  title = {Fermionic Monte Carlo Study of a Realistic Model of Twisted Bilayer Graphene},
  author = {Hofmann, Johannes S. and Khalaf, Eslam and Vishwanath, Ashvin and Berg, Erez and Lee, Jong Yeon},
  journal = {Phys. Rev. X},
  volume = {12},
  issue = {1},
  pages = {011061},
  numpages = {32},
  year = {2022},
  month = {Mar},
  publisher = {American Physical Society},
  doi = {10.1103/PhysRevX.12.011061},
  url = {https://link.aps.org/doi/10.1103/PhysRevX.12.011061}
}

@article{calderon20,
  title = {Interactions in the 8-orbital model for twisted bilayer graphene},
  author = {Calder\'on, M. J. and Bascones, E.},
  journal = {Phys. Rev. B},
  volume = {102},
  issue = {15},
  pages = {155149},
  numpages = {9},
  year = {2020},
  month = {Oct},
  publisher = {American Physical Society},
  doi = {10.1103/PhysRevB.102.155149},
  url = {https://link.aps.org/doi/10.1103/PhysRevB.102.155149}
}

@article{Pizarro19,
  title = {Internal screening and dielectric engineering in magic-angle twisted bilayer graphene},
  author = {Pizarro, J. M. and R\"osner, M. and Thomale, R. and Valent\'{\i}, R. and Wehling, T. O.},
  journal = {Phys. Rev. B},
  volume = {100},
  issue = {16},
  pages = {161102},
  numpages = {6},
  year = {2019},
  month = {Oct},
  publisher = {American Physical Society},
  doi = {10.1103/PhysRevB.100.161102},
  url = {https://link.aps.org/doi/10.1103/PhysRevB.100.161102}
}

@article{Stepanov2020,
  title = {Untying the insulating and superconducting orders in magic-angle graphene},
  volume = {583},
  ISSN = {1476-4687},
  url = {http://dx.doi.org/10.1038/s41586-020-2459-6},
  DOI = {10.1038/s41586-020-2459-6},
  number = {7816},
  journal = {Nature},
  publisher = {Springer Science and Business Media LLC},
  author = {Stepanov,  Petr and Das,  Ipsita and Lu,  Xiaobo and Fahimniya,  Ali and Watanabe,  Kenji and Taniguchi,  Takashi and Koppens,  Frank H. L. and Lischner,  Johannes and Levitov,  Leonid and Efetov,  Dmitri K.},
  year = {2020},
  month = jul,
  pages = {375–378}
}

@article{zhou24,
  title = {Kondo phase in twisted bilayer graphene},
  author = {Zhou, Geng-Dong and Wang, Yi-Jie and Tong, Ninghua and Song, Zhi-Da},
  journal = {Phys. Rev. B},
  volume = {109},
  issue = {4},
  pages = {045419},
  numpages = {26},
  year = {2024},
  month = {Jan},
  publisher = {American Physical Society},
  doi = {10.1103/PhysRevB.109.045419},
  url = {https://link.aps.org/doi/10.1103/PhysRevB.109.045419}
}

@article{Uri2020,
  title = {Mapping the twist-angle disorder and Landau levels in magic-angle graphene},
  volume = {581},
  ISSN = {1476-4687},
  url = {http://dx.doi.org/10.1038/s41586-020-2255-3},
  DOI = {10.1038/s41586-020-2255-3},
  number = {7806},
  journal = {Nature},
  publisher = {Springer Science and Business Media LLC},
  author = {Uri,  A. and Grover,  S. and Cao,  Y. and Crosse,  J. A. and Bagani,  K. and Rodan-Legrain,  D. and Myasoedov,  Y. and Watanabe,  K. and Taniguchi,  T. and Moon,  P. and Koshino,  M. and Jarillo-Herrero,  P. and Zeldov,  E.},
  year = {2020},
  month = may,
  pages = {47–52}
}

@article{Dutta2025,
  title = {Electric Field-Tunable Superconductivity with Competing Orders in Twisted Bilayer Graphene near the Magic Angle},
  volume = {19},
  ISSN = {1936-086X},
  url = {http://dx.doi.org/10.1021/acsnano.4c12770},
  DOI = {10.1021/acsnano.4c12770},
  number = {5},
  journal = {ACS Nano},
  publisher = {American Chemical Society (ACS)},
  author = {Dutta,  Ranit and Ghosh,  Ayan and Mandal,  Shinjan and Watanabe,  Kenji and Taniguchi,  Takashi and Krishnamurthy,  H. R. and Banerjee,  Sumilan and Jain,  Manish and Das,  Anindya},
  year = {2025},
  month = feb,
  pages = {5353–5362}
}

@article{Long2024,
  title = {Evolution of superconductivity in twisted graphene multilayers},
  volume = {121},
  ISSN = {1091-6490},
  url = {http://dx.doi.org/10.1073/pnas.2405259121},
  DOI = {10.1073/pnas.2405259121},
  number = {32},
  journal = {Proceedings of the National Academy of Sciences},
  publisher = {Proceedings of the National Academy of Sciences},
  author = {Long,  Min and Jimeno-Pozo,  Alejandro and Sainz-Cruz,  Héctor and Pantaleón,  Pierre A. and Guinea,  Francisco},
  year = {2024},
  month = jul 
}

@article{putzer25,
  title = {Eliashberg theory and superfluid stiffness of band-off-diagonal pairing in twisted graphene},
  author = {Putzer, Bernhard and Scheurer, Mathias S.},
  journal = {Phys. Rev. B},
  volume = {111},
  issue = {14},
  pages = {144513},
  numpages = {16},
  year = {2025},
  month = {Apr},
  publisher = {American Physical Society},
  doi = {10.1103/PhysRevB.111.144513},
  url = {https://link.aps.org/doi/10.1103/PhysRevB.111.144513}
}

@article{Christos2023,
  title = {Nodal band-off-diagonal superconductivity in twisted graphene superlattices},
  volume = {14},
  ISSN = {2041-1723},
  url = {http://dx.doi.org/10.1038/s41467-023-42471-4},
  DOI = {10.1038/s41467-023-42471-4},
  number = {1},
  journal = {Nature Communications},
  publisher = {Springer Science and Business Media LLC},
  author = {Christos,  Maine and Sachdev,  Subir and Scheurer,  Mathias S.},
  year = {2023},
  month = nov 
}

@article{Yankowitz2019,
  title = {Tuning superconductivity in twisted bilayer graphene},
  volume = {363},
  ISSN = {1095-9203},
  url = {http://dx.doi.org/10.1126/science.aav1910},
  DOI = {10.1126/science.aav1910},
  number = {6431},
  journal = {Science},
  publisher = {American Association for the Advancement of Science (AAAS)},
  author = {Yankowitz,  Matthew and Chen,  Shaowen and Polshyn,  Hryhoriy and Zhang,  Yuxuan and Watanabe,  K. and Taniguchi,  T. and Graf,  David and Young,  Andrea F. and Dean,  Cory R.},
  year = {2019},
  month = mar,
  pages = {1059–1064}
}

@misc{ingham2023quadraticdiracfermionscompetition,
      title={Quadratic Dirac fermions and the competition of ordered states in twisted bilayer graphene}, 
      author={Julian Ingham and Tommy Li and Mathias S. Scheurer and Harley D. Scammell},
      year={2023},
      eprint={2308.00748},
      archivePrefix={arXiv},
      primaryClass={cond-mat.str-el},
      url={https://arxiv.org/abs/2308.00748}, 
}

@article{chichinadze20,
  title = {Nematic superconductivity in twisted bilayer graphene},
  author = {Chichinadze, Dmitry V. and Classen, Laura and Chubukov, Andrey V.},
  journal = {Phys. Rev. B},
  volume = {101},
  issue = {22},
  pages = {224513},
  numpages = {18},
  year = {2020},
  month = {Jun},
  publisher = {American Physical Society},
  doi = {10.1103/PhysRevB.101.224513},
  url = {https://link.aps.org/doi/10.1103/PhysRevB.101.224513}
}

@article{Wu18,
  title = {Theory of Phonon-Mediated Superconductivity in Twisted Bilayer Graphene},
  author = {Wu, Fengcheng and MacDonald, A. H. and Martin, Ivar},
  journal = {Phys. Rev. Lett.},
  volume = {121},
  issue = {25},
  pages = {257001},
  numpages = {6},
  year = {2018},
  month = {Dec},
  publisher = {American Physical Society},
  doi = {10.1103/PhysRevLett.121.257001},
  url = {https://link.aps.org/doi/10.1103/PhysRevLett.121.257001}
}

@article{lian19,
  title = {Twisted Bilayer Graphene: A Phonon-Driven Superconductor},
  author = {Lian, Biao and Wang, Zhijun and Bernevig, B. Andrei},
  journal = {Phys. Rev. Lett.},
  volume = {122},
  issue = {25},
  pages = {257002},
  numpages = {6},
  year = {2019},
  month = {Jun},
  publisher = {American Physical Society},
  doi = {10.1103/PhysRevLett.122.257002},
  url = {https://link.aps.org/doi/10.1103/PhysRevLett.122.257002}
}

@article{Wu19,
  title = {Phonon-induced giant linear-in-$T$ resistivity in magic angle twisted bilayer graphene: Ordinary strangeness and exotic superconductivity},
  author = {Wu, Fengcheng and Hwang, Euyheon and Das Sarma, Sankar},
  journal = {Phys. Rev. B},
  volume = {99},
  issue = {16},
  pages = {165112},
  numpages = {13},
  year = {2019},
  month = {Apr},
  publisher = {American Physical Society},
  doi = {10.1103/PhysRevB.99.165112},
  url = {https://link.aps.org/doi/10.1103/PhysRevB.99.165112}
}

@article{Chou24,
  title = {Topological flat bands, valley polarization, and interband superconductivity in magic-angle twisted bilayer graphene with proximitized spin-orbit couplings},
  author = {Chou, Yang-Zhi and Tan, Yuting and Wu, Fengcheng and Das Sarma, Sankar},
  journal = {Phys. Rev. B},
  volume = {110},
  issue = {4},
  pages = {L041108},
  numpages = {8},
  year = {2024},
  month = {Jul},
  publisher = {American Physical Society},
  doi = {10.1103/PhysRevB.110.L041108},
  url = {https://link.aps.org/doi/10.1103/PhysRevB.110.L041108}
}

@article{Roy19,
  title = {Unconventional superconductivity in nearly flat bands in twisted bilayer graphene},
  author = {Roy, Bitan and Juri\ifmmode \check{c}\else \v{c}\fi{}i\ifmmode \acute{c}\else \'{c}\fi{}, Vladimir},
  journal = {Phys. Rev. B},
  volume = {99},
  issue = {12},
  pages = {121407},
  numpages = {5},
  year = {2019},
  month = {Mar},
  publisher = {American Physical Society},
  doi = {10.1103/PhysRevB.99.121407},
  url = {https://link.aps.org/doi/10.1103/PhysRevB.99.121407}
}

@article{Dong23,
  title = {Superconductivity near spin and valley orders in graphene multilayers},
  author = {Dong, Zhiyu and Levitov, Leonid and Chubukov, Andrey V.},
  journal = {Phys. Rev. B},
  volume = {108},
  issue = {13},
  pages = {134503},
  numpages = {12},
  year = {2023},
  month = {Oct},
  publisher = {American Physical Society},
  doi = {10.1103/PhysRevB.108.134503},
  url = {https://link.aps.org/doi/10.1103/PhysRevB.108.134503}
}

@article{Khalaf2021,
  title = {Charged skyrmions and topological origin of superconductivity in magic-angle graphene},
  volume = {7},
  ISSN = {2375-2548},
  url = {http://dx.doi.org/10.1126/sciadv.abf5299},
  DOI = {10.1126/sciadv.abf5299},
  number = {19},
  journal = {Science Advances},
  publisher = {American Association for the Advancement of Science (AAAS)},
  author = {Khalaf,  Eslam and Chatterjee,  Shubhayu and Bultinck,  Nick and Zaletel,  Michael P. and Vishwanath,  Ashvin},
  year = {2021},
  month = may 
}

@article{Morissette2023,
  title = {Dirac revivals drive a resonance response in twisted bilayer graphene},
  volume = {19},
  ISSN = {1745-2481},
  url = {http://dx.doi.org/10.1038/s41567-023-02060-0},
  DOI = {10.1038/s41567-023-02060-0},
  number = {8},
  journal = {Nature Physics},
  publisher = {Springer Science and Business Media LLC},
  author = {Morissette,  Erin and Lin,  Jiang-Xiazi and Sun,  Dihao and Zhang,  Liangji and Liu,  Song and Rhodes,  Daniel and Watanabe,  Kenji and Taniguchi,  Takashi and Hone,  James and Pollanen,  Johannes and Scheurer,  Mathias S. and Lilly,  Michael and Mounce,  Andrew and Li,  J. I. A.},
  year = {2023},
  month = may,
  pages = {1156–1162}
}

@article{Arora2020,
  title = {Superconductivity in metallic twisted bilayer graphene stabilized by WSe2},
  volume = {583},
  ISSN = {1476-4687},
  url = {http://dx.doi.org/10.1038/s41586-020-2473-8},
  DOI = {10.1038/s41586-020-2473-8},
  number = {7816},
  journal = {Nature},
  publisher = {Springer Science and Business Media LLC},
  author = {Arora,  Harpreet Singh and Polski,  Robert and Zhang,  Yiran and Thomson,  Alex and Choi,  Youngjoon and Kim,  Hyunjin and Lin,  Zhong and Wilson,  Ilham Zaky and Xu,  Xiaodong and Chu,  Jiun-Haw and Watanabe,  Kenji and Taniguchi,  Takashi and Alicea,  Jason and Nadj-Perge,  Stevan},
  year = {2020},
  month = jul,
  pages = {379–384}
}

@article{DiBattista2022,
  title = {Revealing the Thermal Properties of Superconducting Magic-Angle Twisted Bilayer Graphene},
  volume = {22},
  ISSN = {1530-6992},
  url = {http://dx.doi.org/10.1021/acs.nanolett.1c04512},
  DOI = {10.1021/acs.nanolett.1c04512},
  number = {16},
  journal = {Nano Letters},
  publisher = {American Chemical Society (ACS)},
  author = {Di Battista,  Giorgio and Seifert,  Paul and Watanabe,  Kenji and Taniguchi,  Takashi and Fong,  Kin Chung and Principi,  Alessandro and Efetov,  Dmitri K.},
  year = {2022},
  month = aug,
  pages = {6465–6470}
}

@article{Cao2021,
  title = {Nematicity and competing orders in superconducting magic-angle graphene},
  volume = {372},
  ISSN = {1095-9203},
  url = {http://dx.doi.org/10.1126/science.abc2836},
  DOI = {10.1126/science.abc2836},
  number = {6539},
  journal = {Science},
  publisher = {American Association for the Advancement of Science (AAAS)},
  author = {Cao,  Yuan and Rodan-Legrain,  Daniel and Park,  Jeong Min and Yuan,  Noah F. Q. and Watanabe,  Kenji and Taniguchi,  Takashi and Fernandes,  Rafael M. and Fu,  Liang and Jarillo-Herrero,  Pablo},
  year = {2021},
  month = apr,
  pages = {264–271}
}

@article{wang24_2,
  title = {Molecular Pairing in Twisted Bilayer Graphene Superconductivity},
  author = {Wang, Yi-Jie and Zhou, Geng-Dong and Peng, Shi-Yu and Lian, Biao and Song, Zhi-Da},
  journal = {Phys. Rev. Lett.},
  volume = {133},
  issue = {14},
  pages = {146001},
  numpages = {9},
  year = {2024},
  month = {Sep},
  publisher = {American Physical Society},
  doi = {10.1103/PhysRevLett.133.146001},
  url = {https://link.aps.org/doi/10.1103/PhysRevLett.133.146001}
}

@article{Saito2020,
  title = {Independent superconductors and correlated insulators in twisted bilayer graphene},
  volume = {16},
  ISSN = {1745-2481},
  url = {http://dx.doi.org/10.1038/s41567-020-0928-3},
  DOI = {10.1038/s41567-020-0928-3},
  number = {9},
  journal = {Nature Physics},
  publisher = {Springer Science and Business Media LLC},
  author = {Saito,  Yu and Ge,  Jingyuan and Watanabe,  Kenji and Taniguchi,  Takashi and Young,  Andrea F.},
  year = {2020},
  month = jun,
  pages = {926–930}
}

@article{sanchez24_2,
  title = {Nematic versus Kekul\'e Phases in Twisted Bilayer Graphene under Hydrostatic Pressure},
  author = {S\'anchez S\'anchez, Miguel and D\'{\i}az, Israel and Gonz\'alez, Jos\'e and Stauber, Tobias},
  journal = {Phys. Rev. Lett.},
  volume = {133},
  issue = {26},
  pages = {266603},
  numpages = {7},
  year = {2024},
  month = {Dec},
  publisher = {American Physical Society},
  doi = {10.1103/PhysRevLett.133.266603},
  url = {https://link.aps.org/doi/10.1103/PhysRevLett.133.266603}
}

@article{lopez20,
  title = {Electrical band flattening, valley flux, and superconductivity in twisted trilayer graphene},
  author = {Lopez-Bezanilla, Alejandro and Lado, J. L.},
  journal = {Phys. Rev. Res.},
  volume = {2},
  issue = {3},
  pages = {033357},
  numpages = {10},
  year = {2020},
  month = {Sep},
  publisher = {American Physical Society},
  doi = {10.1103/PhysRevResearch.2.033357},
  url = {https://link.aps.org/doi/10.1103/PhysRevResearch.2.033357}
}

@article{dossantos12,
  title = {Continuum model of the twisted graphene bilayer},
  author = {Lopes dos Santos, J. M. B. and Peres, N. M. R. and Castro Neto, A. H.},
  journal = {Phys. Rev. B},
  volume = {86},
  issue = {15},
  pages = {155449},
  numpages = {12},
  year = {2012},
  month = {Oct},
  publisher = {American Physical Society},
  doi = {10.1103/PhysRevB.86.155449},
  url = {https://link.aps.org/doi/10.1103/PhysRevB.86.155449}
}

@article{nam17,
  title = {Lattice relaxation and energy band modulation in twisted bilayer graphene},
  author = {Nam, Nguyen N. T. and Koshino, Mikito},
  journal = {Phys. Rev. B},
  volume = {96},
  issue = {7},
  pages = {075311},
  numpages = {12},
  year = {2017},
  month = {Aug},
  publisher = {American Physical Society},
  doi = {10.1103/PhysRevB.96.075311},
  url = {https://link.aps.org/doi/10.1103/PhysRevB.96.075311}
}

@article{laissandiere10,
author = {Trambly de Laissardière, G. and Mayou, D. and Magaud, L.},
title = {Localization of Dirac Electrons in Rotated Graphene Bilayers},
journal = {Nano Letters},
volume = {10},
number = {3},
pages = {804-808},
year = {2010},
doi = {10.1021/nl902948m},
    note ={PMID: 20121163},
URL = { 
     https://doi.org/10.1021/nl902948m
},
eprint = {     
        https://doi.org/10.1021/nl902948m
}
}

@article{moon12,
  title = {Energy spectrum and quantum Hall effect in twisted bilayer graphene},
  author = {Moon, Pilkyung and Koshino, Mikito},
  journal = {Phys. Rev. B},
  volume = {85},
  issue = {19},
  pages = {195458},
  numpages = {9},
  year = {2012},
  month = {May},
  publisher = {American Physical Society},
  doi = {10.1103/PhysRevB.85.195458},
  url = {https://link.aps.org/doi/10.1103/PhysRevB.85.195458}
}

@article{zhu24,
  title = {Weak-coupling theory of magic-angle twisted bilayer graphene},
  author = {Zhu, Jihang and Torre, Iacopo and Polini, Marco and MacDonald, A. H.},
  journal = {Phys. Rev. B},
  volume = {110},
  issue = {12},
  pages = {L121117},
  numpages = {8},
  year = {2024},
  month = {Sep},
  publisher = {American Physical Society},
  doi = {10.1103/PhysRevB.110.L121117},
  url = {https://link.aps.org/doi/10.1103/PhysRevB.110.L121117}
}

@misc{peng2025manybodyperturbationtheorymoire,
      title={Many-body perturbation theory for moir\'{e} systems}, 
      author={Liangtao Peng and Giovanni Vignale and Shaffique Adam},
      year={2025},
      eprint={2502.06968},
      archivePrefix={arXiv},
      primaryClass={cond-mat.str-el},
      url={https://arxiv.org/abs/2502.06968}, 
}

@misc{xiao2025interactingenergybandsmagic,
      title={The Interacting Energy Bands of Magic Angle Twisted Bilayer Graphene Revealed by the Quantum Twisting Microscope}, 
      author={J. Xiao and A. Inbar and J. Birkbeck and N. Gershon and Y. Zamir and T. Taniguchi and K. Watanabe and E. Berg and S. Ilani},
      year={2025},
      eprint={2506.20738},
      archivePrefix={arXiv},
      primaryClass={cond-mat.mes-hall},
      url={https://arxiv.org/abs/2506.20738}, 
}

@article{Xie2019,
  title = {Spectroscopic signatures of many-body correlations in magic-angle twisted bilayer graphene},
  volume = {572},
  ISSN = {1476-4687},
  url = {http://dx.doi.org/10.1038/s41586-019-1422-x},
  DOI = {10.1038/s41586-019-1422-x},
  number = {7767},
  journal = {Nature},
  publisher = {Springer Science and Business Media LLC},
  author = {Xie,  Yonglong and Lian,  Biao and J\"{a}ck,  Berthold and Liu,  Xiaomeng and Chiu,  Cheng-Li and Watanabe,  Kenji and Taniguchi,  Takashi and Bernevig,  B. Andrei and Yazdani,  Ali},
  year = {2019},
  month = jul,
  pages = {101–105}
}

@article{Wong2020,
  title = {Cascade of electronic transitions in magic-angle twisted bilayer graphene},
  volume = {582},
  ISSN = {1476-4687},
  url = {http://dx.doi.org/10.1038/s41586-020-2339-0},
  DOI = {10.1038/s41586-020-2339-0},
  number = {7811},
  journal = {Nature},
  publisher = {Springer Science and Business Media LLC},
  author = {Wong,  Dillon and Nuckolls,  Kevin P. and Oh,  Myungchul and Lian,  Biao and Xie,  Yonglong and Jeon,  Sangjun and Watanabe,  Kenji and Taniguchi,  Takashi and Bernevig,  B. Andrei and Yazdani,  Ali},
  year = {2020},
  month = jun,
  pages = {198–202}
}

@article{Jiang2019,
  title = {Charge order and broken rotational symmetry in magic-angle twisted bilayer graphene},
  volume = {573},
  ISSN = {1476-4687},
  url = {http://dx.doi.org/10.1038/s41586-019-1460-4},
  DOI = {10.1038/s41586-019-1460-4},
  number = {7772},
  journal = {Nature},
  publisher = {Springer Science and Business Media LLC},
  author = {Jiang,  Yuhang and Lai,  Xinyuan and Watanabe,  Kenji and Taniguchi,  Takashi and Haule,  Kristjan and Mao,  Jinhai and Andrei,  Eva Y.},
  year = {2019},
  month = jul,
  pages = {91–95}
}

@article{carr19_2,
  title = {Derivation of Wannier orbitals and minimal-basis tight-binding Hamiltonians for twisted bilayer graphene: First-principles approach},
  author = {Carr, Stephen and Fang, Shiang and Po, Hoi Chun and Vishwanath, Ashvin and Kaxiras, Efthimios},
  journal = {Phys. Rev. Res.},
  volume = {1},
  issue = {3},
  pages = {033072},
  numpages = {11},
  year = {2019},
  month = {Nov},
  publisher = {American Physical Society},
  doi = {10.1103/PhysRevResearch.1.033072},
  url = {https://link.aps.org/doi/10.1103/PhysRevResearch.1.033072}
}

@article{koshino18,
  title = {Maximally Localized Wannier Orbitals and the Extended Hubbard Model for Twisted Bilayer Graphene},
  author = {Koshino, Mikito and Yuan, Noah F. Q. and Koretsune, Takashi and Ochi, Masayuki and Kuroki, Kazuhiko and Fu, Liang},
  journal = {Phys. Rev. X},
  volume = {8},
  issue = {3},
  pages = {031087},
  numpages = {12},
  year = {2018},
  month = {Sep},
  publisher = {American Physical Society},
  doi = {10.1103/PhysRevX.8.031087},
  url = {https://link.aps.org/doi/10.1103/PhysRevX.8.031087}
}

@article{Saito2021,
  title = {Isospin Pomeranchuk effect in twisted bilayer graphene},
  volume = {592},
  ISSN = {1476-4687},
  url = {http://dx.doi.org/10.1038/s41586-021-03409-2},
  DOI = {10.1038/s41586-021-03409-2},
  number = {7853},
  journal = {Nature},
  publisher = {Springer Science and Business Media LLC},
  author = {Saito,  Yu and Yang,  Fangyuan and Ge,  Jingyuan and Liu,  Xiaoxue and Taniguchi,  Takashi and Watanabe,  Kenji and Li,  J. I. A. and Berg,  Erez and Young,  Andrea F.},
  year = {2021},
  month = apr,
  pages = {220–224}
}

@article{Rozen2021,
  title = {Entropic evidence for a Pomeranchuk effect in magic-angle graphene},
  volume = {592},
  ISSN = {1476-4687},
  url = {http://dx.doi.org/10.1038/s41586-021-03319-3},
  DOI = {10.1038/s41586-021-03319-3},
  number = {7853},
  journal = {Nature},
  publisher = {Springer Science and Business Media LLC},
  author = {Rozen,  Asaf and Park,  Jeong Min and Zondiner,  Uri and Cao,  Yuan and Rodan-Legrain,  Daniel and Taniguchi,  Takashi and Watanabe,  Kenji and Oreg,  Yuval and Stern,  Ady and Berg,  Erez and Jarillo-Herrero,  Pablo and Ilani,  Shahal},
  year = {2021},
  month = apr,
  pages = {214–219}
}

@article{faulstich23,
  title = {Interacting models for twisted bilayer graphene: A quantum chemistry approach},
  author = {Faulstich, Fabian M. and Stubbs, Kevin D. and Zhu, Qinyi and Soejima, Tomohiro and Dilip, Rohit and Zhai, Huanchen and Kim, Raehyun and Zaletel, Michael P. and Chan, Garnet Kin-Lic and Lin, Lin},
  journal = {Phys. Rev. B},
  volume = {107},
  issue = {23},
  pages = {235123},
  numpages = {18},
  year = {2023},
  month = {Jun},
  publisher = {American Physical Society},
  doi = {10.1103/PhysRevB.107.235123},
  url = {https://link.aps.org/doi/10.1103/PhysRevB.107.235123}
}

@article{xie20,
  title = {Nature of the Correlated Insulator States in Twisted Bilayer Graphene},
  author = {Xie, Ming and MacDonald, A. H.},
  journal = {Phys. Rev. Lett.},
  volume = {124},
  issue = {9},
  pages = {097601},
  numpages = {6},
  year = {2020},
  month = {Mar},
  publisher = {American Physical Society},
  doi = {10.1103/PhysRevLett.124.097601},
  url = {https://link.aps.org/doi/10.1103/PhysRevLett.124.097601}
}

@misc{zang2022realspacerepresentationtopological,
      title={Real space representation of topological system: twisted bilayer graphene as an example}, 
      author={Jiawei Zang and Jie Wang and Antoine Georges and Jennifer Cano and Andrew J. Millis},
      year={2022},
      eprint={2210.11573},
      archivePrefix={arXiv},
      primaryClass={cond-mat.mes-hall},
      url={https://arxiv.org/abs/2210.11573}, 
}

@article{song21,
  title = {Twisted bilayer graphene. II. Stable symmetry anomaly},
  author = {Song, Zhi-Da and Lian, Biao and Regnault, Nicolas and Bernevig, B. Andrei},
  journal = {Phys. Rev. B},
  volume = {103},
  issue = {20},
  pages = {205412},
  numpages = {18},
  year = {2021},
  month = {May},
  publisher = {American Physical Society},
  doi = {10.1103/PhysRevB.103.205412},
  url = {https://link.aps.org/doi/10.1103/PhysRevB.103.205412}
}

@article{gonzalez21,
  title = {Magnetic phases from competing Hubbard and extended Coulomb interactions in twisted bilayer graphene},
  author = {Gonz\'alez, J. and Stauber, T.},
  journal = {Phys. Rev. B},
  volume = {104},
  issue = {11},
  pages = {115110},
  numpages = {9},
  year = {2021},
  month = {Sep},
  publisher = {American Physical Society},
  doi = {10.1103/PhysRevB.104.115110},
  url = {https://link.aps.org/doi/10.1103/PhysRevB.104.115110}
}

@article{adhikari24,
  title = {Strongly interacting phases in twisted bilayer graphene at the magic angle},
  author = {Adhikari, Khagendra and Seo, Kangjun and Beach, K. S. D. and Uchoa, Bruno},
  journal = {Phys. Rev. B},
  volume = {110},
  issue = {12},
  pages = {L121123},
  numpages = {7},
  year = {2024},
  month = {Sep},
  publisher = {American Physical Society},
  doi = {10.1103/PhysRevB.110.L121123},
  url = {https://link.aps.org/doi/10.1103/PhysRevB.110.L121123}
}

@article{colomes18,
  title = {Antichiral Edge States in a Modified Haldane Nanoribbon},
  author = {Colom\'es, E. and Franz, M.},
  journal = {Phys. Rev. Lett.},
  volume = {120},
  issue = {8},
  pages = {086603},
  numpages = {5},
  year = {2018},
  month = {Feb},
  publisher = {American Physical Society},
  doi = {10.1103/PhysRevLett.120.086603},
  url = {https://link.aps.org/doi/10.1103/PhysRevLett.120.086603}
}

@article{soejima20,
  title = {Efficient simulation of moir\'e materials using the density matrix renormalization group},
  author = {Soejima, Tomohiro and Parker, Daniel E. and Bultinck, Nick and Hauschild, Johannes and Zaletel, Michael P.},
  journal = {Phys. Rev. B},
  volume = {102},
  issue = {20},
  pages = {205111},
  numpages = {26},
  year = {2020},
  month = {Nov},
  publisher = {American Physical Society},
  doi = {10.1103/PhysRevB.102.205111},
  url = {https://link.aps.org/doi/10.1103/PhysRevB.102.205111}
}

@article{Romanova2022,
  title = {Stochastic many-body calculations of moiré states in twisted bilayer graphene at high pressures},
  volume = {8},
  ISSN = {2057-3960},
  url = {http://dx.doi.org/10.1038/s41524-022-00697-8},
  DOI = {10.1038/s41524-022-00697-8},
  number = {1},
  journal = {npj Computational Materials},
  publisher = {Springer Science and Business Media LLC},
  author = {Romanova,  Mariya and Vlček,  Vojtěch},
  year = {2022},
  month = jan 
}

@article{potasz21,
  title = {Exact Diagonalization for Magic-Angle Twisted Bilayer Graphene},
  author = {Potasz, Pawel and Xie, Ming and MacDonald, A. H.},
  journal = {Phys. Rev. Lett.},
  volume = {127},
  issue = {14},
  pages = {147203},
  numpages = {6},
  year = {2021},
  month = {Sep},
  publisher = {American Physical Society},
  doi = {10.1103/PhysRevLett.127.147203},
  url = {https://link.aps.org/doi/10.1103/PhysRevLett.127.147203}
}

@article{vanhala20,
  title = {Constrained random phase approximation of the effective Coulomb interaction in lattice models of twisted bilayer graphene},
  author = {Vanhala, Tuomas I. and Pollet, Lode},
  journal = {Phys. Rev. B},
  volume = {102},
  issue = {3},
  pages = {035154},
  numpages = {17},
  year = {2020},
  month = {Jul},
  publisher = {American Physical Society},
  doi = {10.1103/PhysRevB.102.035154},
  url = {https://link.aps.org/doi/10.1103/PhysRevB.102.035154}
}

@article{lian21,
  title = {Twisted bilayer graphene. IV. Exact insulator ground states and phase diagram},
  author = {Lian, Biao and Song, Zhi-Da and Regnault, Nicolas and Efetov, Dmitri K. and Yazdani, Ali and Bernevig, B. Andrei},
  journal = {Phys. Rev. B},
  volume = {103},
  issue = {20},
  pages = {205414},
  numpages = {41},
  year = {2021},
  month = {May},
  publisher = {American Physical Society},
  doi = {10.1103/PhysRevB.103.205414},
  url = {https://link.aps.org/doi/10.1103/PhysRevB.103.205414}
}

@article{bernevig21,
  title = {Twisted bilayer graphene. III. Interacting Hamiltonian and exact symmetries},
  author = {Bernevig, B. Andrei and Song, Zhi-Da and Regnault, Nicolas and Lian, Biao},
  journal = {Phys. Rev. B},
  volume = {103},
  issue = {20},
  pages = {205413},
  numpages = {34},
  year = {2021},
  month = {May},
  publisher = {American Physical Society},
  doi = {10.1103/PhysRevB.103.205413},
  url = {https://link.aps.org/doi/10.1103/PhysRevB.103.205413}
}

@article{bernevig21_2,
  title = {Twisted bilayer graphene. V. Exact analytic many-body excitations in Coulomb Hamiltonians: Charge gap, Goldstone modes, and absence of Cooper pairing},
  author = {Bernevig, B. Andrei and Lian, Biao and Cowsik, Aditya and Xie, Fang and Regnault, Nicolas and Song, Zhi-Da},
  journal = {Phys. Rev. B},
  volume = {103},
  issue = {20},
  pages = {205415},
  numpages = {39},
  year = {2021},
  month = {May},
  publisher = {American Physical Society},
  doi = {10.1103/PhysRevB.103.205415},
  url = {https://link.aps.org/doi/10.1103/PhysRevB.103.205415}
}

@misc{zhang2025heavyfermionsmassrenormalization,
      title={Heavy fermions, mass renormalization and local moments in magic-angle twisted bilayer graphene via planar tunneling spectroscopy}, 
      author={Zhenyuan Zhang and Shuang Wu and Dumitru Călugăru and Haoyu Hu and Takashi Taniguchi and Kenji Wanatabe and Andrei B. Bernevig and Eva Y. Andrei},
      year={2025},
      eprint={2503.17875},
      archivePrefix={arXiv},
      primaryClass={cond-mat.mes-hall},
      url={https://arxiv.org/abs/2503.17875}, 
}

@article{Wu2021,
  title = {Chern insulators,  van Hove singularities and topological flat bands in magic-angle twisted bilayer graphene},
  volume = {20},
  ISSN = {1476-4660},
  url = {http://dx.doi.org/10.1038/s41563-020-00911-2},
  DOI = {10.1038/s41563-020-00911-2},
  number = {4},
  journal = {Nature Materials},
  publisher = {Springer Science and Business Media LLC},
  author = {Wu,  Shuang and Zhang,  Zhenyuan and Watanabe,  K. and Taniguchi,  T. and Andrei,  Eva Y.},
  year = {2021},
  month = feb,
  pages = {488–494}
}

@article{Datta2023,
  title = {Heavy quasiparticles and cascades without symmetry breaking in twisted bilayer graphene},
  volume = {14},
  ISSN = {2041-1723},
  url = {http://dx.doi.org/10.1038/s41467-023-40754-4},
  DOI = {10.1038/s41467-023-40754-4},
  number = {1},
  journal = {Nature Communications},
  publisher = {Springer Science and Business Media LLC},
  author = {Datta,  Anushree and Calderón,  M. J. and Camjayi,  A. and Bascones,  E.},
  year = {2023},
  month = aug 
}

@article{rai24,
  title = {Dynamical Correlations and Order in Magic-Angle Twisted Bilayer Graphene},
  author = {Rai, Gautam and Crippa, Lorenzo and C\ifmmode \u{a}\else \u{a}\fi{}lug\ifmmode \u{a}\else \u{a}\fi{}ru, Dumitru and Hu, Haoyu and Paoletti, Francesca and de' Medici, Luca and Georges, Antoine and Bernevig, B. Andrei and Valent\'{\i}, Roser and Sangiovanni, Giorgio and Wehling, Tim},
  journal = {Phys. Rev. X},
  volume = {14},
  issue = {3},
  pages = {031045},
  numpages = {22},
  year = {2024},
  month = {Sep},
  publisher = {American Physical Society},
  doi = {10.1103/PhysRevX.14.031045},
  url = {https://link.aps.org/doi/10.1103/PhysRevX.14.031045}
}

@article{Nuckolls2023,
  title = {Quantum textures of the many-body wavefunctions in magic-angle graphene},
  volume = {620},
  ISSN = {1476-4687},
  url = {http://dx.doi.org/10.1038/s41586-023-06226-x},
  DOI = {10.1038/s41586-023-06226-x},
  number = {7974},
  journal = {Nature},
  publisher = {Springer Science and Business Media LLC},
  author = {Nuckolls,  Kevin P. and Lee,  Ryan L. and Oh,  Myungchul and Wong,  Dillon and Soejima,  Tomohiro and Hong,  Jung Pyo and Călugăru,  Dumitru and Herzog-Arbeitman,  Jonah and Bernevig,  B. Andrei and Watanabe,  Kenji and Taniguchi,  Takashi and Regnault,  Nicolas and Zaletel,  Michael P. and Yazdani,  Ali},
  year = {2023},
  month = aug,
  pages = {525–532}
}

@article{Nuckolls2020,
  title = {Strongly correlated Chern insulators in magic-angle twisted bilayer graphene},
  volume = {588},
  ISSN = {1476-4687},
  url = {http://dx.doi.org/10.1038/s41586-020-3028-8},
  DOI = {10.1038/s41586-020-3028-8},
  number = {7839},
  journal = {Nature},
  publisher = {Springer Science and Business Media LLC},
  author = {Nuckolls,  Kevin P. and Oh,  Myungchul and Wong,  Dillon and Lian,  Biao and Watanabe,  Kenji and Taniguchi,  Takashi and Bernevig,  B. Andrei and Yazdani,  Ali},
  year = {2020},
  month = dec,
  pages = {610–615}
}

@article{Chen2024,
  title = {Strong electron–phonon coupling in magic-angle twisted bilayer graphene},
  volume = {636},
  ISSN = {1476-4687},
  url = {http://dx.doi.org/10.1038/s41586-024-08227-w},
  DOI = {10.1038/s41586-024-08227-w},
  number = {8042},
  journal = {Nature},
  publisher = {Springer Science and Business Media LLC},
  author = {Chen,  Cheng and Nuckolls,  Kevin P. and Ding,  Shuhan and Miao,  Wangqian and Wong,  Dillon and Oh,  Myungchul and Lee,  Ryan L. and He,  Shanmei and Peng,  Cheng and Pei,  Ding and Li,  Yiwei and Hao,  Chenyue and Yan,  Haoran and Xiao,  Hanbo and Gao,  Han and Li,  Qiao and Zhang,  Shihao and Liu,  Jianpeng and He,  Lin and Watanabe,  Kenji and Taniguchi,  Takashi and Jozwiak,  Chris and Bostwick,  Aaron and Rotenberg,  Eli and Li,  Chu and Han,  Xu and Pan,  Ding and Liu,  Zhongkai and Dai,  Xi and Liu,  Chaoxing and Bernevig,  B. Andrei and Wang,  Yao and Yazdani,  Ali and Chen,  Yulin},
  year = {2024},
  month = dec,
  pages = {342–347}
}

@article{angeli19,
  title = {Valley Jahn-Teller Effect in Twisted Bilayer Graphene},
  author = {Angeli, M. and Tosatti, E. and Fabrizio, M.},
  journal = {Phys. Rev. X},
  volume = {9},
  issue = {4},
  pages = {041010},
  numpages = {17},
  year = {2019},
  month = {Oct},
  publisher = {American Physical Society},
  doi = {10.1103/PhysRevX.9.041010},
  url = {https://link.aps.org/doi/10.1103/PhysRevX.9.041010}
}

@misc{sánchez2025fermivelocitymagicangle,
      title={Fermi velocity and magic angle renormalization in twisted bilayer graphene}, 
      author={Miguel Sánchez Sánchez and José González and Tobias Stauber},
      year={2025},
      eprint={2508.12825},
      archivePrefix={arXiv},
      primaryClass={cond-mat.mes-hall},
      url={https://arxiv.org/abs/2508.12825}, 
}

@article{zhao25,
  title = {Topological Mott localization and pseudogap metal in twisted bilayer graphene},
  author = {Zhao, Jing-Yu and Zhou, Boran and Zhang, Ya-Hui},
  journal = {Phys. Rev. B},
  volume = {112},
  issue = {8},
  pages = {085107},
  numpages = {25},
  year = {2025},
  month = {Aug},
  publisher = {American Physical Society},
  doi = {10.1103/9n8v-7rx2},
  url = {https://link.aps.org/doi/10.1103/9n8v-7rx2}
}

@article{Kerelsky2019,
  title = {Maximized electron interactions at the magic angle in twisted bilayer graphene},
  volume = {572},
  ISSN = {1476-4687},
  url = {http://dx.doi.org/10.1038/s41586-019-1431-9},
  DOI = {10.1038/s41586-019-1431-9},
  number = {7767},
  journal = {Nature},
  publisher = {Springer Science and Business Media LLC},
  author = {Kerelsky,  Alexander and McGilly,  Leo J. and Kennes,  Dante M. and Xian,  Lede and Yankowitz,  Matthew and Chen,  Shaowen and Watanabe,  K. and Taniguchi,  T. and Hone,  James and Dean,  Cory and Rubio,  Angel and Pasupathy,  Abhay N.},
  year = {2019},
  month = jul,
  pages = {95–100}
}

@article{Lu2019,
  title = {Superconductors,  orbital magnets and correlated states in magic-angle bilayer graphene},
  volume = {574},
  ISSN = {1476-4687},
  url = {http://dx.doi.org/10.1038/s41586-019-1695-0},
  DOI = {10.1038/s41586-019-1695-0},
  number = {7780},
  journal = {Nature},
  publisher = {Springer Science and Business Media LLC},
  author = {Lu,  Xiaobo and Stepanov,  Petr and Yang,  Wei and Xie,  Ming and Aamir,  Mohammed Ali and Das,  Ipsita and Urgell,  Carles and Watanabe,  Kenji and Taniguchi,  Takashi and Zhang,  Guangyu and Bachtold,  Adrian and MacDonald,  Allan H. and Efetov,  Dmitri K.},
  year = {2019},
  month = oct,
  pages = {653–657}
}


\appendix

\setcounter{figure}{0}
\renewcommand\thefigure{\thesection\arabic{figure}}    

\setcounter{table}{0}
\renewcommand\thetable{\thesection\arabic{table}}

\section{Lattice geometry and tight-binding Hamiltonian}{\label{appa}}

\begin{figure}[t]
    \centering
    \includegraphics[width=.7\linewidth]{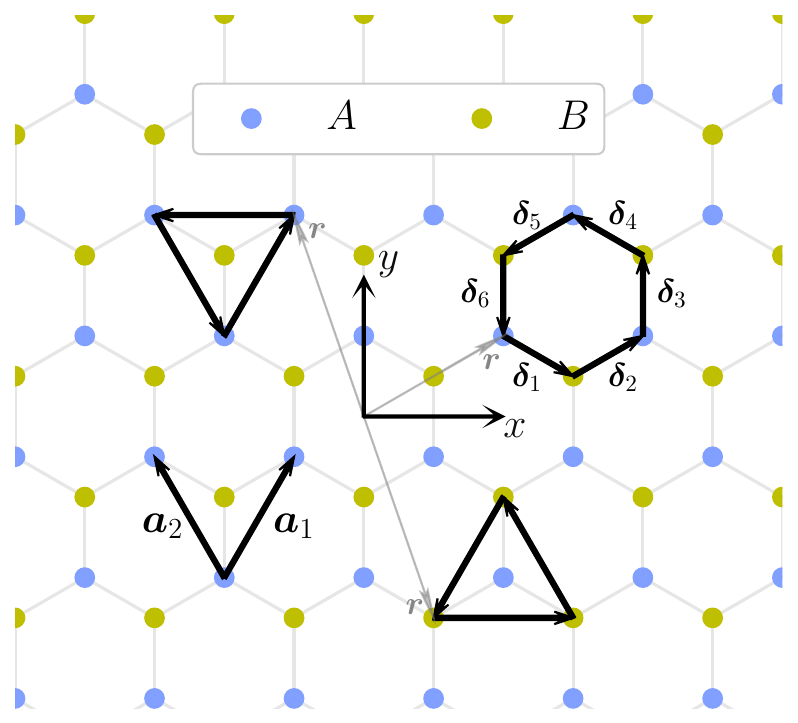}
    \caption{Lattice of monolayer graphene. We label the $A$ and $B$ sublattices, the unit vectors ($\boldsymbol{a}_1,\boldsymbol{a}_2$), the nearest neighbors vectors ($\boldsymbol{\delta}_i$) and the triangular and hexagonal loops used to compute the wave function overlaps.}
    \label{lattice}
\end{figure}

A layer of graphene has the two basis vectors $\a_1=a_0(\sqrt{3}/2,3/2)$ and $\a_2=a_0(-\sqrt{3}/2,3/2)$ with $a_0=1.42$\r{A}. We label the set of atoms displaced by $(\a_1 + \a_2)/3$ from the lattice positions as the $A$ sublattice, and those displaced by $-(\a_1 + \a_2)/3$, as the $B$ sublattice, see Fig. \ref{lattice}.

In twisted bilayer graphene, the top and bottom layers are located at the equilibrium vertical distances $z=\pm d_0/2 = \pm 1.6755$ $\text{\AA}$. The top layer gets rotated by an angle $\theta/2$ and the bottom layer gets rotated by $-\theta/2$ around the $z$ axis. In this work we choose the twist angle $\theta = 1.0501 ^\circ = \cos^{-1}(1-1/(6n^2+6n+2))$ with $n=31$, giving a commensurate superlattice \cite{dossantos12}. The moiré lattice constant is $L_M=13.4$ nm and the number of atoms in the unit cell is $11908$. The unit vectors of the moiré lattice are $\boldsymbol{L}_1 = L_M(\sqrt{3}/2,1/2)$ and $\boldsymbol{L}_2 = L_M(-\sqrt{3}/2,1/2)$, and the moiré reciprocal lattice has the unit vectors $\boldsymbol{g}_1 = 4\pi/\sqrt{3}L_M(1/2,\sqrt{3}/2)$ and $\boldsymbol{g}_2 = 4\pi/\sqrt{3}L_M(-1/2,\sqrt{3}/2)$. We have included the effects of in-plane lattice relaxation using the theory of Ref. \cite{nam17}.

The atomistic Hamiltonian consists of the tight-binding part $H_\text{TB}$ and the interaction part $H_{\text{int}}$,
\begin{align}
    & {H} =  H_{\text{TB}} + {H}_\text{int},  \\ 
    & H_{\text{TB}} = \sum_{\boldsymbol{r}\boldsymbol{r'}s} t(\boldsymbol{r}-\boldsymbol{r'}) c^\dagger_{\boldsymbol{r}s}c_{\boldsymbol{r'}s},   \\  
    & {H}_\text{int} = \frac{1}{2}\sum_{\boldsymbol{r}\neq \boldsymbol{r'},ss'} V(\boldsymbol{r}- \boldsymbol{r'}) \delta n_{\boldsymbol{r}s}   \delta n_{\boldsymbol{r'}s'} + \frac{1}{2}\sum_{\boldsymbol{r}s} U  \delta n_{\boldsymbol{r}s}  \delta n_{\boldsymbol{r}\Bar{s}},
    \label{hamiltoniansm}
\end{align}
where $c^\dagger_{\boldsymbol{r}s} (c_{\boldsymbol{r}s})$ creates (annihilates) an electron with spin $s$ at site $\boldsymbol{r}$, $\delta n_{\boldsymbol{r}s} = c^\dagger_{\boldsymbol{r}s} c_{\boldsymbol{r}s} - 1/2$ is the spin $s$ density at site $\boldsymbol{r}$ relative to $1/2$ and $\Bar s$ is the spin opposite to $s$. The hopping function $t(\rr)$ is given by \cite{laissandiere10,moon12}
\begin{align}
    &t(\boldsymbol{r}) = - V_{pp\pi}(r) \big(1 - \cos^2(\varphi_{\boldsymbol{r}} )\big) + V_{pp\sigma}(r) \cos^2(\varphi_{\boldsymbol{r}}), \\
    &V_{pp\pi}(r) = V_{pp \pi}^0 \ \text{exp}\big(-(r - a_{cc})/r_0\big), \\
    &V_{pp\sigma}(r) = V_{pp \sigma}^0 \ \text{exp}\big(-(r - d_0)/r_0\big),
\end{align}
with  $\boldsymbol{r}=(x,y,z)$, $r=\sqrt{x^2+y^2+z^2}$, $\cos(\varphi_{\boldsymbol{r}}) = z/r$, $V^0_{ppp\pi} = 2.7 \text{ eV}$, $V^0_{pp\sigma} = 0.48 \text{ eV}$, $a_{cc} = a_0/\sqrt{3}$, $r_0 = 0.0453  \text{ nm}$, and the interaction $V(\rr)$ is the double-gated potential
\begin{align}
 V(\boldsymbol{r}-\boldsymbol{r'}) &= \frac{e^2}{4\pi \epsilon_0 \epsilon_r}\sum_{n=-\infty}^\infty \frac{(-1)^n}{||\boldsymbol{r} - \boldsymbol{r'} + n\xi \boldsymbol{\hat{z}}||} \nonumber \\ &= \frac{14.4 \text{ eV}\times\text{\AA}}{\epsilon_r}\sum_{n=-\infty}^\infty \frac{(-1)^n}{||\boldsymbol{r} - \boldsymbol{r'} + 100 \text{\AA} \times n \boldsymbol{\hat{z}} ||},
\end{align}
with $\epsilon_r=10$ being the effective dielectric constant and $\xi=10$ nm the distance between the metallic gates, located at $z=\pm \xi/2$. $U=4$ eV is the on-site Hubbard energy. 

This Hamiltonian is solved at the charge neutrality point in mean-field (Hartree-Fock) theory. We obtain the symmetric state by enforcing the crystallographic symmetries $C_{2z}$ and $C_{3z}$ and the time-reversal symmetry, $\mathcal{T}$, at each step of the self-consistency loop. For the symmetry-broken states, we input the self-consistent solutions of the projected theory (Ref. \cite{sanchez25}) as initial seeds of the algorithm. In order to avoid converging to a lower-energy solution, we enforce $C_{2z}$ and $\mathcal{T}$ for the NSM and TIVC states, $C_{2z}\mathcal{T}$ for the VP state and $C_{2z}$ for the QAH state. For a detailed account of our self-consistent Hartree-Fock method and the many-body projection algorithm for obtaining the flat-band projected theory, we refer the reader to Appendix B of Ref. \cite{sanchez25}.

\section{Derivation of the wave function overlaps}{\label{appb}}

For later use, let us denote the graphene nearest neighbors vectors around an hexagonal loop by $\boldsymbol{\delta_1} = -\boldsymbol{\delta_4} = (\boldsymbol{\a_1}-2\boldsymbol{\a_2})/3$, $\boldsymbol{\delta_2} = -\boldsymbol{\delta_5} = (2\boldsymbol{\a_1}-\boldsymbol{\a_2})/3$ and  $\boldsymbol{\delta_3} = -\boldsymbol{\delta_6} = (\boldsymbol{\a_1}+\boldsymbol{\a_2})/3$, see Fig. \ref{lattice}. Also, let us remind the reader that the two $K$ points of graphene are located at $\boldsymbol{K}= (\boldsymbol{G}_2-\boldsymbol{G}_1)/3 = (-4\pi/3\sqrt{3}a_0,0)$ and $\boldsymbol{K'} = -\boldsymbol{K}$, with $\boldsymbol{G}_1=\frac{2\pi}{3a_0}(1,\sqrt{3})$ and $\boldsymbol{G}_2=\frac{2\pi}{3a_0}(-1,\sqrt{3})$ being the unit vectors of the graphene reciprocal lattice. After the twist, the $K$ points of the top and bottom layer rotate to $\boldsymbol{K}_{t} = R_{\theta/2} (\boldsymbol{K}) = - \boldsymbol{K}'_{t}$ and  $\boldsymbol{K}_{b} = R_{- \theta/2} (\boldsymbol{K}) = - \boldsymbol{K}'_{b}$, respectively, with $R_\varphi$ being the rotation by angle $\varphi$.  

The wave function at low energies can be written as the product of a rapidly oscillating phase, due to the graphene $K$ point, times a slow envelope function. A general wave function contains contributions from both valleys,
\begin{align}
|\psi\rangle =& \sum_\rr \psi(\rr) |\rr \rangle  =\sum_{\sigma \eta \ell} \sum_{\rr \in  \sigma \ell} e^{i \eta \K_\ell \cdot\rr}f_{\sigma \eta \ell}(\rr) |\rr \rangle
\label{wavefunction}
\end{align}

Here $\sigma=A,B$ is the graphene sublattice, $\eta=1(K), -1(K')$ the graphene valley, and $\ell=t(\text{top}), b(\text{bottom})$ the layer; $\rr \in \sigma \ell$ denotes the sum over atoms on sublattice $\sigma$ and layer $\ell$. With lattice relaxation, the localized states $|\rr \rangle$ are centered at the relaxed atomic positions, but we choose to write the wave function coefficients as functions of the original, unrelaxed, positions (in particular, the phase  $e^{i \eta \K_\ell \cdot\rr}$ is the same as without relaxation) .

The envelope functions $f_{\sigma \eta \ell}(\boldsymbol{r})$ depend on the sublattice, valley and layer. Their slow variation allows the approximation
\begin{align}
f_{\sigma \eta \ell}(\rr+\b) \approx f_{\sigma \eta \ell}(\rr)    
\label{approximation}
\end{align}
for any vector $\b$ with a magnitude of the order of the lattice constant. Moreover, if $|\psi \rangle$ is a Bloch function with momentum $\kk$, we have the modified Bloch condition $f_{\sigma \eta t(b)}(\rr + \boldsymbol{L}_{1,2}) = e^{-(+)2\pi i/3} e^{i \kk \cdot \boldsymbol{L}_{1,2}}f_{\sigma \eta t(b)}(\rr )$, owing to the identity $e^{-i \eta  \boldsymbol{K}_{t(b)} \cdot \boldsymbol{L}_{1}} = e^{-i \eta  \boldsymbol{K}_{t(b)} \cdot \boldsymbol{L}_{2}} = e^{-(+) 2\pi i \eta/3} \neq 1$. In continuum theories, the envelope functions $f_{\sigma \eta \ell}$ are promoted to functions of continuous space, subject to the normalization condition $\sum_{\eta \sigma \ell} \int d\rr |f_{\eta \sigma \ell}(\rr)|^2 = 1$.

In the following we omit the layer index in the envelope functions, as the calculations are identical in both layers. Moreover, for each layer we can work in the frame of reference where the $K$ point and lattice vectors take their unrotated values $\boldsymbol{K}, \a_1$ and $\a_2$.

\begin{widetext}
\paragraph*{Intersublattice, intervalley overlaps.}

Consider the hexagonal loop in Fig. \ref{lattice}, computing the order parameter
\begin{align}
    h(\boldsymbol{r}) =& \psi^*(\boldsymbol{r})\psi(\boldsymbol{r} + \boldsymbol{\delta}_1) + \psi^*(\boldsymbol{r}+\boldsymbol{\delta}_1)\psi(\boldsymbol{r} + \a_1-\a_2) + \psi^*(\boldsymbol{r} + \a_1-\a_2)\psi(\boldsymbol{r} + \a_1-\a_2 + \boldsymbol{\delta}_3) \nonumber \\ &+ \psi^*(\boldsymbol{r} + \a_1-\a_2 + \boldsymbol{\delta}_3) \psi (\boldsymbol{r} + \a_1) + \psi^*(\boldsymbol{r} + \a_1) \psi (\boldsymbol{r} + \a_1 + \boldsymbol{\delta}_5) + \psi^*(\boldsymbol{r} + \a_1 + \boldsymbol{\delta}_5) \psi (\boldsymbol{r}). 
\end{align}
After expanding the wave function according to Eq. (\ref{wavefunction}), and using Eq. (\ref{approximation}), we obtain
\begin{align}
    h(\boldsymbol{r}) =& 3e^{-2i \boldsymbol{K}\cdot \boldsymbol{r}} e^{2\pi i/3} f_{AK}^*(\boldsymbol{r})f_{BK'}(\boldsymbol{r}) + 3e^{-2i \boldsymbol{K}\cdot \boldsymbol{r}} f_{BK}^*(\boldsymbol{r})f_{AK'}(\boldsymbol{r}) + 3e^{2i \boldsymbol{K}\cdot \boldsymbol{r}} f_{BK'}^*(\boldsymbol{r})f_{AK}(\boldsymbol{r}) \nonumber \\
    & + 3e^{2i \boldsymbol{K}\cdot \boldsymbol{r}} e^{-2\pi i/3} f_{AK'}^*(\boldsymbol{r})f_{BK}(\boldsymbol{r})
    \label{deltainrasubintraval}
\end{align}
(the intravalley products $|f_{A,BK}(\boldsymbol{r})|^2$ vanish due to the fact that $e^{i\boldsymbol{K}\cdot \boldsymbol{\delta_1}} + e^{i\boldsymbol{K}\cdot \boldsymbol{\delta_3}} + e^{i\boldsymbol{K}\cdot \boldsymbol{\delta_5}} = 0$). Now, consider 
\begin{align}
h^+(\boldsymbol{r}) = \big(h(\boldsymbol{r}+\a_1) + h(\boldsymbol{r}+\a_2-\a_1) + h(\boldsymbol{r}-\a_2)\big)/3, \\
h^-(\boldsymbol{r}) = \big(h(\boldsymbol{r}-\a_1) + h(\boldsymbol{r}-\a_2+\a_1) + h(\boldsymbol{r}+\a_2)\big)/3.     
\end{align}
Again, from Eqs. \ref{wavefunction} and \ref{approximation}, we have
\begin{align}
    h_+(\boldsymbol{r}) =& 3e^{-2i \boldsymbol{K}\cdot \boldsymbol{r}} f_{AK}^*(\boldsymbol{r})f_{BK'}(\boldsymbol{r}) + 3e^{-2i \boldsymbol{K}\cdot \boldsymbol{r}} e^{-2\pi i /3} f_{BK}^*(\boldsymbol{r})f_{AK'}(\boldsymbol{r}) 
    + 3e^{2i \boldsymbol{K}\cdot \boldsymbol{r}} e^{2 \pi i/3}f_{BK'}^*(\boldsymbol{r})f_{AK}(\boldsymbol{r}) \nonumber \\ &+ 3e^{2i \boldsymbol{K}\cdot \boldsymbol{r}} f_{AK'}^*(\boldsymbol{r})f_{BK}(\boldsymbol{r}),  \\
    h_-(\boldsymbol{r}) =& 3e^{-2i \boldsymbol{K}\cdot \boldsymbol{r}} e^{-2\pi i/3} f_{AK}^*(\boldsymbol{r})f_{BK'}(\boldsymbol{r}) + 3e^{-2i \boldsymbol{K}\cdot \boldsymbol{r}} e^{2\pi i /3} f_{BK}^*(\boldsymbol{r})f_{AK'}(\boldsymbol{r}) 
    + 3e^{2i \boldsymbol{K}\cdot \boldsymbol{r}} e^{-2 \pi i/3}f_{BK'}^*(\boldsymbol{r})f_{AK}(\boldsymbol{r}) \nonumber \\ &+ 3e^{2i \boldsymbol{K}\cdot \boldsymbol{r}} e^{2\pi i/3} f_{AK'}^*(\boldsymbol{r})f_{BK}(\boldsymbol{r}).
\end{align}
We can then solve for the wave function overlaps $f_{BK'}(\boldsymbol{r}) f_{AK}^*(\boldsymbol{r})  $ and $f_{AK'}(\boldsymbol{r}) f_{BK}^*(\boldsymbol{r}) $, with the expressions

\begin{align}
    f_{BK'}(\boldsymbol{r}) f_{AK}^*(\boldsymbol{r}) &= \frac{-ie^{2i\boldsymbol{K}\cdot \boldsymbol{r}}}{9 \sqrt{3}}\bigg(  \Big( h(\boldsymbol{r})  + e^{2\pi i/3}h_+(\boldsymbol{r})  + e^{-2\pi i/3}h_-(\boldsymbol{r})  \Big) - e^{-2 \pi i/3} \Big(h(\boldsymbol{r})^*  + e^{2\pi i/3}h_+(\boldsymbol{r})^* + e^{-2\pi i/3}h_-(\boldsymbol{r})^* \Big) \bigg),
    \label{overlapintersubinterval1} \\
     f_{AK'}(\boldsymbol{r}) f_{BK}^*(\boldsymbol{r}) &=  \frac{-i e^{2i\boldsymbol{K}\cdot \boldsymbol{r}}}{9\sqrt{3}}\bigg( \Big(h(\boldsymbol{r})^*  + e^{2\pi i/3}h_+(\boldsymbol{r})^*  + e^{-2\pi i/3}h_-(\boldsymbol{r})^*  \Big) - e^{-2 \pi i/3}\Big(h(\boldsymbol{r})  + e^{2\pi i/3}h_+(\boldsymbol{r})  + e^{-2\pi i/3}h_-(\boldsymbol{r})  \Big)  \bigg).
    \label{overlapintersubinterval2}
\end{align}

\paragraph*{Intersublattice, intravalley overlaps.}

We obtain the intersublattice, intravalley overlaps by attaching appropriate phases to the hexagonal loops in Fig. \ref{lattice} (see also Fig. \ref{loops}). We compute the two quantities,
\begin{align}
    h^\omega_{1}(\boldsymbol{r}) =& e^{2\pi i/3}\psi^*(\boldsymbol{r})\psi(\boldsymbol{r} +\boldsymbol{\delta}_1) +  e^{-2\pi i/3}\psi^*(\boldsymbol{r}+\boldsymbol{\delta}_1)\psi(\boldsymbol{r} + \a_1-\a_2) +\psi^*(\boldsymbol{r} + \a_1-\a_2)\psi(\boldsymbol{r} + \a_1-\a_2 + \boldsymbol{\delta}_3) \nonumber \\ &+ e^{2\pi i/3}\psi^*(\boldsymbol{r} + \a_1-\a_2 + \boldsymbol{\delta}_3) \psi (\boldsymbol{r} + \a_1) + e^{-2\pi i/3}\psi^*(\boldsymbol{r} + \a_1) \psi (\boldsymbol{r} + \a_1 + \boldsymbol{\delta}_5) + \psi^*(\boldsymbol{r} + \a_1 + \boldsymbol{\delta}_5) \psi (\boldsymbol{r}), \\
    h^\omega_{2}(\boldsymbol{r}) =& e^{-2\pi i/3}\psi^*(\boldsymbol{r})\psi(\boldsymbol{r} +\boldsymbol{\delta}_1) +  e^{2\pi i/3}\psi^*(\boldsymbol{r}+\boldsymbol{\delta}_1)\psi(\boldsymbol{r} + \a_1-\a_2) + e^{2\pi i/3}\psi^*(\boldsymbol{r} + \a_1-\a_2)\psi(\boldsymbol{r} + \a_1-\a_2 + \boldsymbol{\delta}_3) \nonumber \\ &+ \psi^*(\boldsymbol{r} + \a_1-\a_2 + \boldsymbol{\delta}_3) \psi (\boldsymbol{r} + \a_1) + \psi^*(\boldsymbol{r} + \a_1) \psi (\boldsymbol{r} + \a_1 + \boldsymbol{\delta}_5) + e^{-2\pi i/3} \psi^*(\boldsymbol{r} + \a_1 + \boldsymbol{\delta}_5) \psi (\boldsymbol{r}), 
\end{align}
After expanding $\psi(\boldsymbol{r})$ using Eqs. \ref{wavefunction} and \ref{approximation} like before and performing some algebra, we obtain
\begin{align}
    h^\omega_1(\rr) =& 3 f_{AK}^*(\rr)f_{BK}(\rr) + 3 f_{BK'}^*(\rr)f_{AK'}(\rr) \\
    h^\omega_2(\rr) =& 3e^{2\pi i/3} f_{AK}^*(\rr)f_{BK}(\rr) + 3e^{-2 \pi i/3} f_{BK'}^*(\rr)f_{AK'}(\rr).
\end{align}
We can then solve for the intersublattice, intravalley overlaps,
\begin{align}
    f_{BK}(\rr) f_{AK}^*(\rr) = \frac{-i}{3\sqrt{3}} \bigg( h_2^\omega(\rr) - e^{- 2\pi i/3}h_{1}^\omega(\rr)\bigg), \label{overlapintersubintraval1} \\
    f_{AK'}(\rr) f_{BK'}^*(\rr) = \frac{i}{3\sqrt{3}} \bigg( h_2^\omega(\rr) - e^{2 \pi i/3} h_{1}^\omega(\rr)\bigg). \label{overlapintersubintraval2}
\end{align}

\paragraph*{Intrasublattice, intravalley overlaps.}

The triangular loops in Fig. \ref{lattice} correspond to the following quantities:
\begin{align}
    \Delta_A(\rr) = \psi^*(\rr) \psi(\rr+\a_2-\a_1) + \psi^*(\rr+\a_2-\a_1) \psi(\rr-\a_1) + \psi^*(\rr-\a_1)\psi(\rr), \\
    \Delta_B(\rr) = \psi^*(\rr) \psi(\rr-\a_1-\a_2) + \psi^*(\rr-\a_1-\a_2) \psi(\rr + \a_1) + \psi^*(\rr+\a_1)\psi(\rr), 
\end{align}
for the sublattice $A$ and sublattice $B$ triangles, respectively. Then, we have
\begin{align}
    |f_{AK}(\rr)|^2 - |f_{AK'}(\rr)|^2 = \frac{-2}{3\sqrt{3}} \text{Im} \Delta_A(\rr), \label{overlapintrasubintraval1} \\
    |f_{AK}(\rr)|^2 + |f_{AK'}(\rr)|^2 = \frac{-2}{3} \text{Re} \Delta_A(\rr), \label{overlapintrasubintraval2} \\
    |f_{BK}(\rr)|^2 - |f_{BK'}(\rr)|^2 = \frac{2}{3\sqrt{3}} \text{Im} \Delta_B(\rr), \label{overlapintrasubintraval3} \\
    |f_{BK}(\rr)|^2 + |f_{BK'}(\rr)|^2 = \frac{-2}{3} \text{Re} \Delta_B(\rr). \label{overlapintrasubintraval4}
\end{align}
Hence the triangular loops compute the valley polarizations $|f_{A,BK}(\rr)|^2 - |f_{A,BK'}(\rr)|^2$ \cite{lopez20,colomes18}, and densities $|f_{A,BK}(\rr)|^2 + |f_{A,BK'}(\rr)|^2$, or equivalently, the intrasublattice intravalley products $|f_{A,BK}(\rr)|^2, |f_{A,BK'}(\rr)|^2$.

\paragraph*{Intrasublattice, intervalley overlaps.}

For the intrasublattice, intervalley product we consider again the triangular loops in Fig. \ref{lattice}. We modify the loops with appropriate phases (see Fig. \ref{loops}),
\begin{align}
    \Delta^\omega_{B}(\boldsymbol{r}) = \psi^*(\boldsymbol{r})\psi(\boldsymbol{r} + \a_1 - \a_2)  
    + e^{2\pi i /3}  \psi^*(\boldsymbol{r} + \a_1 - \a_2) \psi(\boldsymbol{r} + \a_1) 
    + e^{- 2\pi i /3}  \psi^*(\boldsymbol{r} + \a_1) \psi(\boldsymbol{r}), \\
    \Delta^\omega_{A}(\boldsymbol{r}) = \psi^*(\boldsymbol{r})\psi(\boldsymbol{r} + \a_2 - \a_1) + e^{-2\pi i/3} \psi^*(\boldsymbol{r}+\a_2 - \a_1)\psi(\boldsymbol{r}-\a_1) + e^{2\pi i/3} \psi^*(\boldsymbol{r}-\a_1)\psi(\boldsymbol{r}),
\end{align}
with $\Delta_{BKK'}(\boldsymbol{r})$ computed from the sublattice $B$ triangles and $\Delta_{BKK'}(\boldsymbol{r})$ computed from the sublattice $A$ triangles. In this case, the intravalley contributions cancel, and after some straightforward algebra we have
\begin{align}
    \Delta^\omega_{B}(\boldsymbol{r}) &= 3e^{2i \boldsymbol{K}\cdot \boldsymbol{r}}e^{2\pi i /3}  f^*_{BK'}(\boldsymbol{r})f_{BK}(\boldsymbol{r}), \\
    \Delta^\omega_{A}(\boldsymbol{r}) &= 3e^{2i \boldsymbol{K}\cdot \boldsymbol{r}}e^{-2\pi i /3}  f^*_{AK'}(\boldsymbol{r})f_{AK}(\boldsymbol{r}),
\end{align}
and wave function overlaps follow immediately,
\begin{align}
    f_{BK}(\boldsymbol{r}) f^*_{BK'}(\boldsymbol{r})  &= e^{-2\pi i /3}e^{-2i \boldsymbol{K}\cdot \boldsymbol{r}} \Delta_{B}^\omega(\boldsymbol{r})/3, \label{overlapintrasubinterval1} \\
     f_{AK}(\boldsymbol{r}) f^*_{AK'}(\boldsymbol{r}) &= e^{2\pi i /3}e^{-2i \boldsymbol{K}\cdot \boldsymbol{r}} \Delta_{A}^\omega(\boldsymbol{r})/3. \label{overlapintrasubinterval2}
\end{align}
\end{widetext}

Let us note now that the intervalley products of a Bloch state are not periodic, and acquire an extra phase upon translation coming from the modified Bloch condition mentioned above.

To conclude, let us mention that in practice one can compute the local order parameters $\sum_{\substack{\text{occupied} \\ \text{states}}} f_{\sigma \eta \ell}(\rr) f_{\sigma'\eta'\ell}^*(\rr)$ of Eq. (\ref{rhomatrix}) directly from the microscopic correlation matrix $\sum_{\substack{\text{occupied} \\ \text{states}}} \psi(\rr) \psi^*(\rr')$, by linearity.
Moreover, for microscopic wave functions normalized as $\sum_\rr |\psi(\rr)|^2=1$, performing the integral over the first unit cell of Eq. (\ref{orderparams}) is equivalent to summing the corresponding microscopic quantities, i.e. the right hand sides of Eqs. 
(\ref{overlapintersubinterval1}), (\ref{overlapintersubinterval2})
(\ref{overlapintersubintraval1}), (\ref{overlapintersubintraval2}), (\ref{overlapintrasubintraval1}), (\ref{overlapintrasubintraval2}), (\ref{overlapintrasubintraval3}), (\ref{overlapintrasubintraval4}), (\ref{overlapintrasubinterval1}), (\ref{overlapintrasubinterval2}), over the lattice points of the unit cell. 

\section{Order parameters as a function of the number of bands}{\label{appc}}

In the main text, we restricted the occupied states in Eq. (\ref{rhomatrix}) to be within the flat bands. Here we report additional results considering the occupied states within the $20$ central bands, i.e. including the first occupied $10$ bands, of the correlated states. By construction, the order parameters in Eq. (\ref{orderparams}) are bounded between $-5$ and $5$ when we include 10 occupied bands; however, we find that their values remain close to those for only the flat bands. This reflects the fact that the symmetry breaking lies predominantly on the flat band manifold. Only $\langle \sigma_0 \tau_0 \mu_0 \rangle$, counting half the number of occupied bands, scales accordingly (we get consistently $\langle \sigma_0 \tau_0 \mu_0 \rangle = 0.997$ for the flat-band case and $4.98$ for the $10$-band case; the small difference from $1$ and $5$ stem from the various approximations of the method).

Even though the decomposition of Eq. (\ref{wavefunction}) is valid only for the states near the Fermi level, one can compute $\rho_{\sigma \eta \ell, \sigma' \eta' \ell} (\rr)$ in Eq. (\ref{rhomatrix}) including all occupied bands, using the full microscopic correlation matrix. One possibility for regularizing the outcome is subtracting the correlation matrix of the symmetric state, i.e. taking the symmetric state as the zero point of the order parameters (in particular, $\langle \sigma_0 \tau_0 \mu_0 \rangle$ then counts half the number of filled bands relative to charge neutrality, and we get $\langle \sigma_0 \tau_0 \mu_0 \rangle = 0.00$ consistently for all states); this way, contributions from the bulk of states deep in the Fermi sea cancel. In fact, we find that to a good approximation $\rho_{\sigma \eta \ell,\sigma' \eta' \ell} = 0$ in the symmetric state, so we get the same values for the order parameters with or without regularization. The properties of $h_{1,2}^\omega(\rr)$ and $\Delta_{A,B}^\omega(\rr)$ under $C_{3z}$ explain the cancellation of the intersublattice, intravalley and intrasublattice, intervalley components, respectively, but the vanishing of the remaining components cannot be explained from symmetry alone. Finally, we notice that the local intersublattice, intervalley orders $\rho_{\sigma \eta \ell,\Bar{\sigma} \Bar{\eta} \ell}(\rr)$ (the bars denote the opposite sublattice/valley) vanish at each $\rr$ to a good approximation, with $\int_{\substack{\text{1º unit } \text{cell}}} d\rr \ |\rho_{\sigma \eta  \ell, \Bar{\sigma} \Bar{\eta} \ell}(\rr)| \sim 10^{-4}$.

In Table \ref{tab1app} we compare the order parameters including the occupied flat bands, the first $10$ occupied bands and all occupied bands. They tend to increase with the number of bands, but the dominant contribution comes from the flat manifold. If we compare the results for the flat bands and the first $10$ bands, the largest variation of the order parameters is of $18\%$ in the OP state. If we include all bands, the variations are larger, reaching up to $51 \%$ in the TIVC state. We stress, however, that the decomposition of Eq. (\ref{wavefunction}) breaks down at high energies, when the electron states are not supported near the Dirac valleys of the graphene sheets. For this reason the interpretation of the order parameters is less clear in this case---even if we have regularized the results using the symmetric state, as described above.

\begin{figure}[H]
    \centering
    \includegraphics[width=0.9\linewidth]{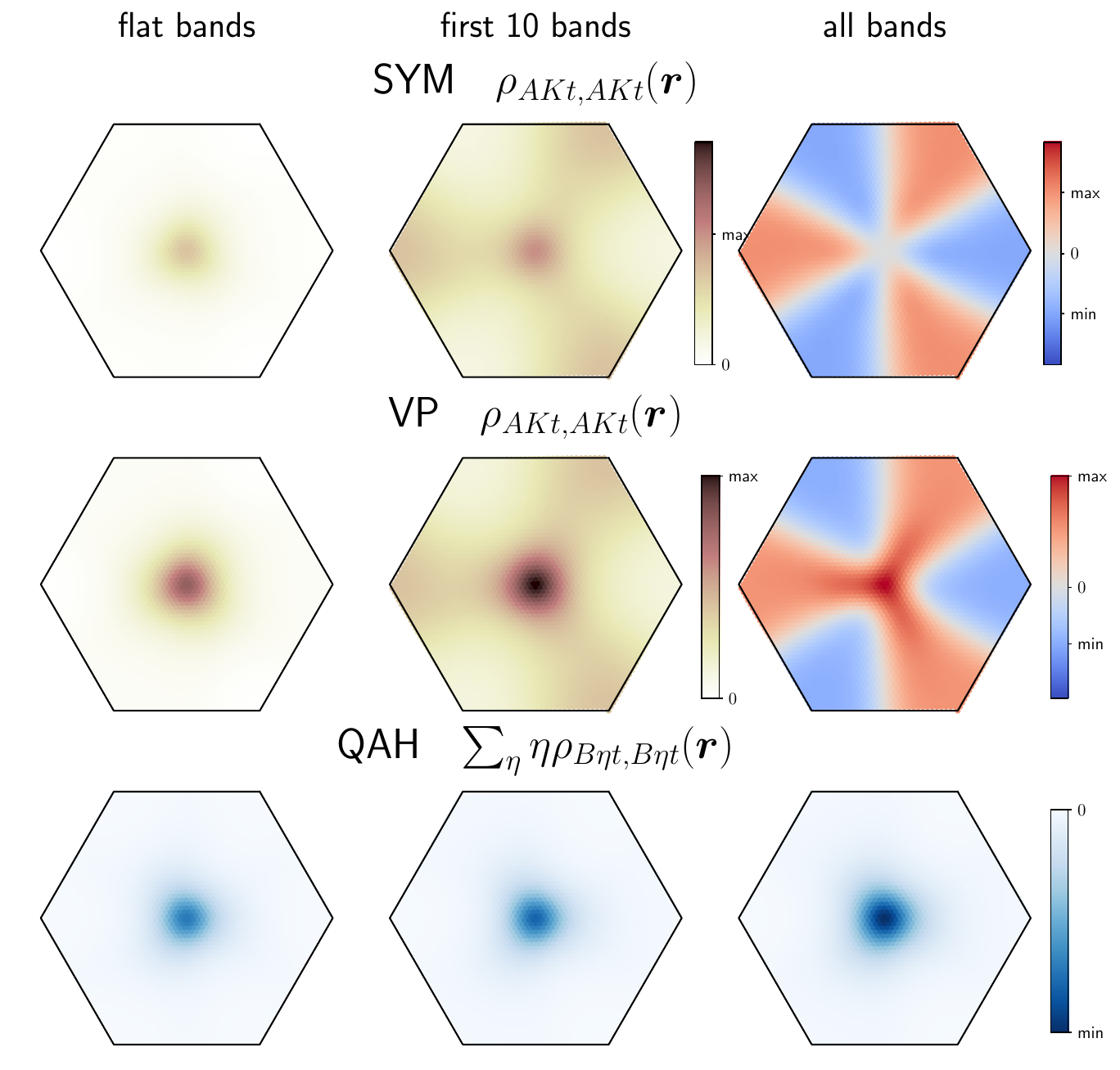}
    \caption{Local order parameters as a function of the number of bands included. The first two rows show the valley $K$ density on the $A$ sublattice $A$ and top layer of the SYM and VP states. The color scales are identical, and the maximal (minimal) values are labeled for each state. The bottom row depicts the valley polarization on the $B$ sublattice and top layer of QAH. All three plots are qualitatively similar.}
    \label{fig2app}
\end{figure}

In Fig. \ref{fig2app} we compare some selected local order parameters including the occupied flat bands, the first $10$ occupied bands, and all bands. We show the valley-$K$ density on the $A$ sublattice and top layer, $\rho_{AKt,AKt}(\rr)$, for the SYM state in the top panel and for the VP state in the middle panel. In the bottom panel we show the valley polarization on $B$ sublattice $B$ and top layer, $\sum_\eta \eta \rho_{B\eta t,B\eta t}(\rr)$, for the QAH state. For the flat and first $10$ bands the valley-resolved density corresponds to the electron density, whereas for all bands the order parameter has positive as well as negative values, and integrates to $0$. Note that when we sum over all bands, the VP values can be interpreted as the background values of the symmetric state plus some excess weight at the center due to the symmetry breaking. On the other hand, by $\mathcal{T}$ symmetry the local valley polarization of the SYM state is exactly $0$. This can be seen in the QAH panel, showing qualitatively similar plots in the three cases.  

TIVC, KIVC, OP and NSM (not shown here) exhibit analogous behaviors. The intersublattice, intervalley order parameters of the TIVC and KIVC are qualitatively similar for the flat, the first $10$ and all bands, which can be understood from the fact that $\rho_{\sigma \eta \ell,\Bar{\sigma} \Bar{\eta} \ell}(\rr) \approx 0$ at each $\rr$ in the normal state. On the other hand, when all bands are included, the intersublattice, intravalley order shows a distribution similar to the normal state. In the NSM,an additional weight appears, similarly to the excess density of the VP state shown in Fig. \ref{fig2app}. This additional weight integrates to the nonzero order parameters of the NSM.\\

\begin{table}[H]
    \centering
    \renewcommand{\arraystretch}{1.25} 
    \begin{tabular}{c c c c c}
        \hline \hline
        &  & flat bands  &  first $10$ bands &  all bands \\
        \hline
        KIVC & $\langle \sigma_y\tau_x\rangle$ & $0.838$ & $0.852$ & $0.905$\\
        \hline
        QAH & $\langle  \sigma_z\tau_z\rangle$ & $0.595$ & $0.704$ & $0.889$\\
        \hline
        OP & $\langle  \sigma_z \rangle$ & $0.590$ & $0.688$ & $0.778$\\
        \hline
        NSM & \begin{tabular}{@{}c@{}} $\langle \sigma_x \mu_z \rangle$ \\ $\langle \sigma_y\tau_z\mu_z\rangle$ \end{tabular} & \begin{tabular}{@{}c@{}} $-0.563$ \\ $0.331$ \end{tabular} &  \begin{tabular}{@{}c@{}} $-0.604$ \\ $0.354$ \end{tabular} & \begin{tabular}{@{}c@{}} $-0.730$ \\ $0.429$ \end{tabular} \\
        \hline
        VP & $\langle \tau_z  \rangle$ & $0.886$ & $ 0.886$ & $0.886$ \\
        \hline
        TIVC & $\langle \sigma_x \tau_x \rangle$ & $-0.615$ & $-0.689$ & $-0.931$ \\
        \hline \hline
    \end{tabular}
    \caption{Order parameters including the occupied flat bands, the first occupied $10$ bands and all occupied bands.}
    \label{tab1app}
\end{table}

\end{document}